\documentclass{pasa}%

\usepackage{graphicx}
\usepackage{tabularx}
\usepackage{multirow}

\title[Magnitudes and $M/L$ ratios of GCs]{Absolute $V$-band magnitudes and mass-to-light ratios of Galactic globular clusters}

\author[Baumgardt et al.]{Baumgardt H.$^1$\thanks{E-mail:
h.baumgardt@uq.edu.au}, Sollima, A.$^2$ and Hilker, M.$^3$\\
$^{1}$ School of Mathematics and Physics, The University of Queensland, St. Lucia, QLD 4072, Australia \\
$^{2}$ INAF Osservatorio Astronomico di Bologna, via Gobetti 93/3, Bologna, 40129, Italy\\ 
$^{3}$ European Southern Observatory, Karl-Schwarzschild-Str. 2, 85748 Garching, Germany\\
}%

\jid{PASA}
\doi{10.1017/pas.\the\year.xxx}
\jyear{\the\year}

\newcommand {\gaia}{\textit{Gaia }}

\usepackage{aas_macros}
\usepackage{hyperref} 
\hypersetup{colorlinks,citecolor=blue,linkcolor=blue,urlcolor=blue}

\begin{document}

\begin{frontmatter}
\maketitle

\begin{abstract}
We have used HST and ground-based photometry to determine total $V$-band magnitudes and mass-to-light ratios of 
more than 150 Galactic globular clusters. We do this by summing up the magnitudes of their individual member stars, using color-magnitude information, 
{\tt Gaia} DR2 proper motions and radial velocities to distinguish cluster stars from background stars. 
Our new magnitudes confirm literature estimates for bright clusters with $V<8$, but can deviate by up to two magnitudes from literature values for fainter clusters.
They lead to absolute mass-to-light ratios that are confined to the narrow range $1.4<M/L_V<2.5$, significantly smaller than what was found before.
We also find a correlation between a cluster's $M/L_V$ value and its age, in agreement with theoretical predictions.
The $M/L_V$ ratios of globular clusters are also in good agreement with those predicted by stellar isochrones, arguing against a significant amount of dark matter 
inside globular clusters. We finally find that, in agreement with what has been seen in M~31, the magnitude distribution of outer halo globular clusters has a tail towards faint clusters 
that is absent in the inner parts of the Milky Way.
\end{abstract}

\end{frontmatter}

\section{INTRODUCTION }
\label{sec:intro}

Globular cluster systems are powerful tools to study the evolution of galaxies since they trace the major star formation episodes of their parent galaxies 
\citep{brodiestrader2006}. Their color and spatial distribution therefore allows to identify different stellar sub-populations, while
their radial velocities can be used to determine the mass profile of galaxies \citep[e.g][]{richtleretal2011,potaetal2015}.
In addition, past merger episodes of galaxies can be deduced from their globular cluster populations \citep{kruijssenetal2020}.

For distant extragalactic globular clusters, one can normally only observe the integrated light of an otherwise unresolved cluster, so
integrated magnitudes and broad-band colors must be used to infer the mass, age and metallicity of each cluster. In order to facilitate
such studies, it is useful to know the same integrated magnitudes and colors of Milky Way globular clusters. This is due to the fact that
Milky Way GCs can be resolved into individual
stars, so that their ages, metallicities and masses can be determined with much higher accuracy through colour-magnitude diagram (CMD) isochrone fitting and high resolution spectroscopy. 

Previous measurements of the total magnitudes of Galactic globular clusters were either based on aperture photometry \citep[e.g.][]{hanesbrodie1985,peterson1986,vanderbeke2014} or 
were derived by integrating the surface brightness profile of a cluster \citep[e.g][]{mclaughlinvandermarel2005}. Both approaches have problems distinguishing
between field and cluster stars. In addition, bright cluster giants are often excluded from the surface density profile, leading to a possible underestimation
of the derived total cluster luminosity. These problems become more severe for bulge clusters that are located in
regions of very high background stellar density, and fainter clusters that contain only few giant stars. 

Recent years have seen a rise in the publication of photometric catalogues presenting deep, HST-based photometry of the centres of globular clusters \citep[e.g.][]{piottoetal2002,sarajedinietal2007},
as well as wide area, ground-based studies \citep[e.g.][]{stetsonetal2019}. In addition, deep photometric data is nowadays also available from ground-based surveys like SDSS \citep{abazajianetal2003},
2MASS \citep{skrutskieetal2006}, DES \citep{despaper2005} or PanStarrs \citep{chambersetal2016}. This makes it possible to determine the cluster magnitudes by summing up the magnitudes of the individual member stars. 
This allows to use the location of stars in a color-magnitude diagram, as well as their proper motions from \gaia DR2 and their radial velocities \citep{baumgardthilker2018} to 
distinguish between cluster and field stars. In addition, \gaia proper motions and photometry can also be used to better determine the density profiles of clusters 
\citep{deboeretal2019}. Both effects can be used to determinate the total cluster magnitude with higher accuracy.

In the present paper we use published photometry to determine new magnitudes and mass-to-light ratios of 153 Galactic globular clusters. We
concentrate on the determination of $V$-band magnitudes since $V$-band data is available for the largest number of clusters. 
Our procedure can however easily be adopted to other wavelength bands. Our paper is organised as follows: In sec.~\ref{sec:phot} we describe the input photometry
used and explain our procedure to derive the total magnitudes. In sec. \ref{sec:results} we compare our magnitudes with published literature values and calculate total magnitudes
and mass-to-light ratios for all clusters. We draw our conclusions in sec. \ref{sec:concl}.

\section{OBSERVATIONAL DATA }
\label{sec:phot}

\subsection{Input Photometry}

%%We use a number of photometric catalogues as input to determine the magnitudes of stars in globular clusters. 
The photometry for the inner parts of globular clusters is mainly
based on Hubble Space Telescope (HST) based observations, since only HST has a sufficiently high spatial resolution to resolve the centers of dense globular clusters. Our 
main source for HST based photometry is the ACS Survey of Galactic Globular Clusters \citep{sarajedinietal2007}.
The ACS Survey has observed the centers of 65 globular clusters using the F606W and F814W filters of the HST ACS/WFC camera. To this set of 65 clusters we add 14 clusters that have been observed
mostly with the HST WFC3 camera in the F438W/F555W filters and were analysed by \citet{baumgardtetal2019}. We furthermore use F439W/F555W WFPC2 photometry from the HST Globular Cluster Snapshot Program \citep{piottoetal2002} 
for 19 globular clusters as well as a number of published literature observations for other clusters. Finally for six globular clusters (AM~4, FSR~1735, NGC~6440, Pal~13, Sagittarius II  and Ter~3)
we downloaded HST images from the {\tt STSci} archive and performed stellar photometry using {\tt DOLPHOT} \citep{dolphin2000, dolphin2016} on it. Photometry was performed on the CTE corrected flc images,
using the point-spread functions provided for each camera and filter combination by {\tt DOLPHOT}. 
Where necessary, we first transformed the HST instrumental coordinates into equatorial coordinates by cross-matching stellar positions and magnitudes from HST with the positions of stars in the \gaia catalogue. 
In total we have been able to obtain deep HST photometry, reaching between two to five magnitudes below the 
main-sequence turn-off for 126 globular clusters. The sources of the used HST photometry are listed in Table~\ref{tab1}. The remaining clusters are mostly low-mass and low-density clusters for which ground-based photometry
should also be sufficiently complete for upper main sequence and giant stars.

Due to the small field-of-view, available HST photometry is largely limited to the innermost 120'' around the centers of globular clusters. For many globular clusters this
is less than the observed half-light radius. We therefore combine the HST photometry in the inner parts with ground-based photometry for the outer cluster parts. Our main source 
for ground-based photometry is the recent catalogue of ground-based photometry by \citet{stetsonetal2019}.
They present wide-field, ground-based photometry in the Johnson-Cousins UBVRI bands based on about 90,000 public and proprietary images for 48 Galactic globular clusters. 
We furthermore use unpublished data that was compiled in a similar way by Peter Stetson and that we downloaded from the Canadian Astronomy Data Centre\footnote{This data is available under
https://www.cadc-ccda.hia-iha.nrc-cnrc.gc.ca/en/community/STETSON/index.html} for an additional 63 clusters. For globular clusters for which $V$-band data by P. Stetson is not available,
we used other ground-based $V$-band data from the literature as indicated in Table~\ref{tab1}. Where necessary, we cross-correlated the ground-based data against the
\gaia catalogue to convert instrumental (x/y) coordinates into (RA/Dec) coordinates.

For two clusters, ESO452-SC11 and IC1257, we performed our own photometry. We downloaded and reduced publicly available data in the V and I bands 
that were taken with EFOSC (mounted on the NTT at La Silla) in May 2012 for ESO~452-SC11 (ESO programme ID: 089.D-0194(A)) and May 2015 for 
IC~1257 (ESO programme ID: 095.D-0037(A)). We used DAOPHOT to perform PSF photometry
on short and long exposures. The photometric calibrations based on colour terms and extinction coefficients provided by ESO for EFOSC, and the zeropoints were
adopted such that they match previous, shallower photometry in the Johnson-Cousins system by \citet{cornishetal2006} for ESO452-SC11 and \citet{harrisetal1997}
for IC1257.

We finally used data from the DECam Plane Survey \citep{schlaflyetal2018} 
%%and the second data release of the Pan-STARRS survey \citep{chambersetal2016} 
for clusters for which we could not find any other photometry. In total, we have been able to obtain deep ground-based photometry 
that covers the giant-branch and turn-over regions for 136 globular clusters. The sources of the ground-based photometry for the individual clusters are listed in Table~\ref{tab1}.

\subsection{Creation of a master catalogue}

Since the HST photometry is not in the standard Johnson-Cousins UBVRI system, we first converted the magnitudes of the various HST camera systems into the Johnson UBVRI system. 
For HST photometry taken from \citet{sarajedinietal2007} and \citet{piottoetal2002}, we use the BVI band magnitudes that were calculated by these authors. For the other data, we apply
a magnitude transformation following \citet{holtzmanetal1995}:
\begin{equation}
 TMAG-SMAG = c_0 + c_1 \times TCOL + c_2 \times TCOL^2 \; ,
\end{equation}
where TMAG is a magnitude in the target system, SMAG is a magnitude in the source system, $c_0$, $c_1$ and $c_2$ are transformation constants, and TCOL is the difference between two magnitudes 
in the target system. Since the right hand side requires a magnitude difference in the target system, we apply the above transformations iteratively, using the color difference in the source system 
as starting value. In order to convert HST/WFPC2 magnitudes to UBVRI magnitudes, we use the coefficients given in Table~7 of \citet{holtzmanetal1995}.
For the transformation of HST/ACS magnitudes we use the transformation coefficients given in Table~18 of \citet{siriannietal2005}, while the conversion of HST WFC3/UVIS magnitudes is done 
using the coefficients given in Table~2 of \citet{harris2018}. 
\begin{figure*}[h]
\begin{center}
\includegraphics[width=0.95\textwidth]{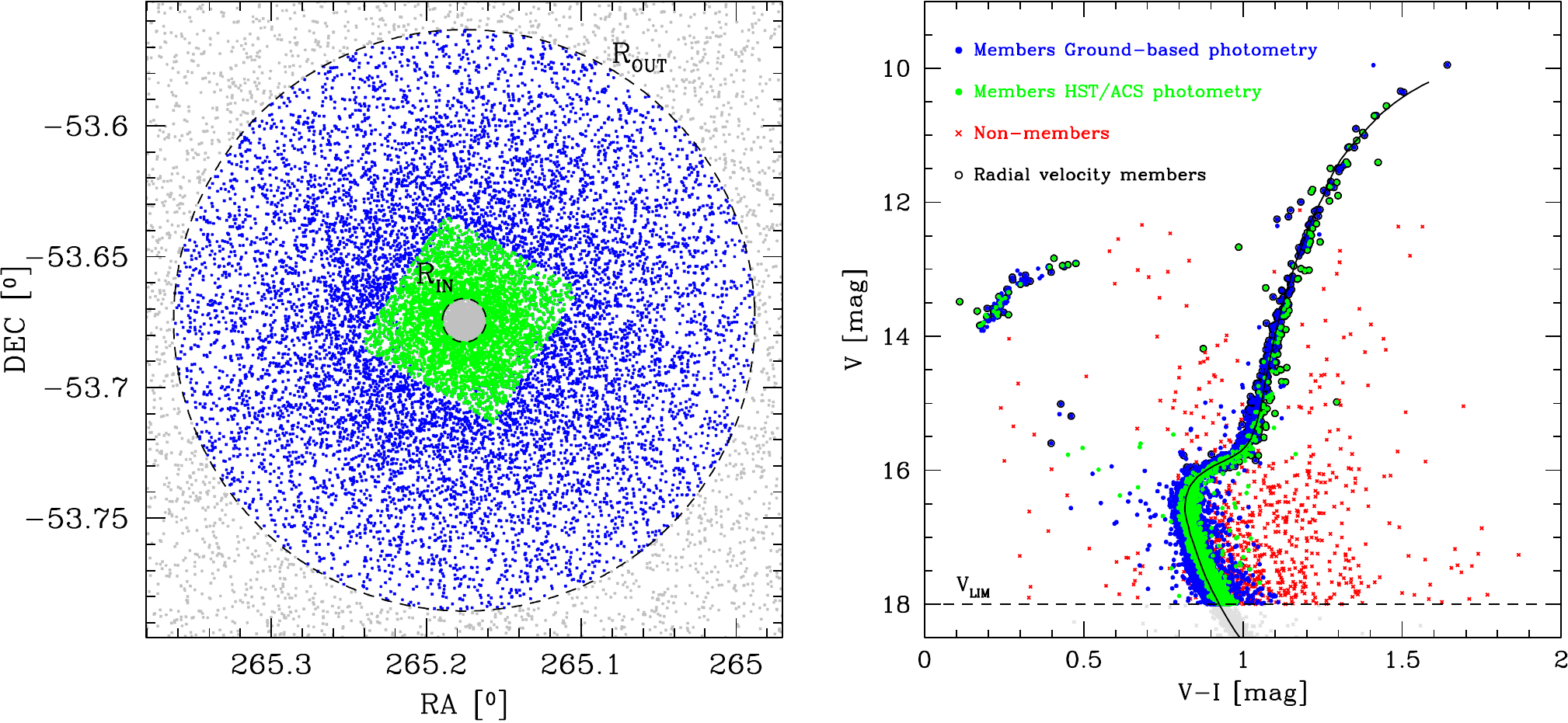}
\caption{Illustration of our member star selection approach for the globular cluster NGC~6397. The left panel shows an 800'' x 800'' arcsec field centered on the cluster. Stars selected from HST/ACS
observations are shown in green, stars from the ground-based photometry of \citet{stetsonetal2019} in blue. The dashed circles show the limits of the field for which we determine the cluster luminosity. The
right panel shows a CMD of NGC~6397 with a 12 Gyr old PARSEC isochrone overlayed as solid line. Blue and green circles depict cluster members while red crosses depict stars classified as non-members based
on their CMD position, Gaia proper motion or radial velocity. The dashed line marks the lower limit down to which we use observed stars. Circles mark radial velocity members.}
%%The lower left panel shows
%%the comparison of the transformed F606W HST magnitudes vs. ground-based V magnitudes for all stars in common after a shift of 0.03 mag has been applied to the HST magnitudes. The lower right panel shows
%%how the estimated V-band magnitude of NGC~6397 changes with increasing $R_{Out}$, with the dashed line marking our final value of 5.44.}
\end{center}
\end{figure*}

To increase the accuracy of the transformed BVI magnitudes that we obtain from the HST photometry, we compare them against ground-based BVI magnitudes for the
stars in common. For the clusters listed in Table~2 that have ground-based photometry, we use the same ground-based photometry for calibration that we also use for the outer cluster parts. 
For three of the remaining clusters without ground-based photometry (NGC~6293, NGC~6304, NGC~6540), we use the photometry from Peter Stetson's standard star archive \citep{stetson2000}.
The resulting magnitude shifts are mostly below 0.05 mag, except for a few heavily reddened bulge clusters for which the corrections can reach 0.2 mag.

After transforming the HST magnitudes to the Johnson-Cousins system, we create a master catalogue for each cluster by combining the HST photometry in the inner parts with the ground-based photometry 
in the outer
cluster parts. We cross-match the positions of stars in the HST catalogue with those from the ground-based photometry using a search radius of 0.5 arcsec. Since the HST photometry has
a higher precision than the ground-based photometry in the crowded cluster centres, we keep the HST photometry for the stars that are in common between both data sets. Fig.~1 
illustrates our member search approach for the cluster NGC~6397.

\subsection{Selection of cluster members}

Cluster members are selected from the photometric master catalogue based on three criteria: Position in the CMD, radial velocity and \gaia proper motion. In order to select
stars based on photometry, we fit PARSEC isochrones \citep{bressanetal2012} to each cluster and use these to select main sequence and giant star members. Possible cluster members must either have
a colour difference no larger than 2.5 times their photometric error from the best-fitting isochrone or have a colour difference less than a maximum value. We choose the maximum color difference 
individually for each cluster based on the observed width of the RGB and the amount of background contamination. For most clusters, these values are usually around 0.20 mag. We also select stars 
as potential cluster members if they are located in the CMD in the region that correspond to horizontal-branch stars and blue stragglers. For a few clusters with strong and variable reddening, we first 
derive a de-reddened CMD by shifting stars along the reddening vector to a common main sequence and then identify cluster members in the de-reddened CMD.
%%by calculating reddening values for each giant star in the cluster from their color difference to the PARSEC isochrone and by then correcting the CMD position of each stars
%%based on the average reddening value of the giant stars around it. Member identification based CMD position is then done for the de-reddened CMD. Fig.~2 gives an examples of reddening corrected
%%CMDs for Pal~2.

Our second criterion for membership determination are the stellar radial velocities compiled by \citet{baumgardt2017} and \citet{baumgardthilker2018}. Their 
data contains radial velocities and membership information for about 250,000 stars in the fields of globular clusters. We cross-match the positions
of all stars that are classified as members based on CMD position with the radial velocities of \citet{baumgardthilker2018} and keep only those stars that either have no radial velocity measurement,
or have a radial velocity that is within $\pm 2.5 \sigma$ of the cluster mean velocity. Here the velocity dispersion $\sigma$ is calculated at the position of each star based on the best-fitting $N$-body 
model of \citet{baumgardthilker2018}.

We finally use the \gaia DR2 proper motions and parallaxes for membership determination. For stars that have passed the CMD and radial velocity tests, we cross-match their positions against the positions
of stars in the \gaia catalogue and require that their proper motion is within $2.5 \sigma$ of the mean cluster proper motion determined by \citet{baumgardtetal2019}. 
We also require that the star has a parallax that is compatible with the cluster parallax $p_{CL}=1/d$ where $d$ is the cluster distance given by \citet{baumgardthilker2018}. 
We keep all stars that
have no \gaia counterparts. Stars without \gaia counterparts are either faint stars, or stars in the centers of clusters that have a high chance of 
being cluster members due to the strong density contrast between cluster and field stars in the center.

\subsection{Magnitude determination}

We calculate the total luminosity of the cluster members determined in the previous section according to:
\begin{equation}
 L_{V Obs} = \sum_i 10^{-0.4 V_i}   
%%-M_{V \odot})}
\end{equation}
%%where $M_{V \odot}=4.83$ is the absolute magnitude of the Sun \citep{binneymerrifield1998}. 
To derive the total cluster luminosities from $L_{V Obs}$, we then need to correct  $L_{V Obs}$ for faint stars not included in the photometry, cluster regions that are not covered by the photometry and
any background contamination remaining in the data.
\begin{figure}[t]
\begin{center}
\includegraphics[width=\columnwidth]{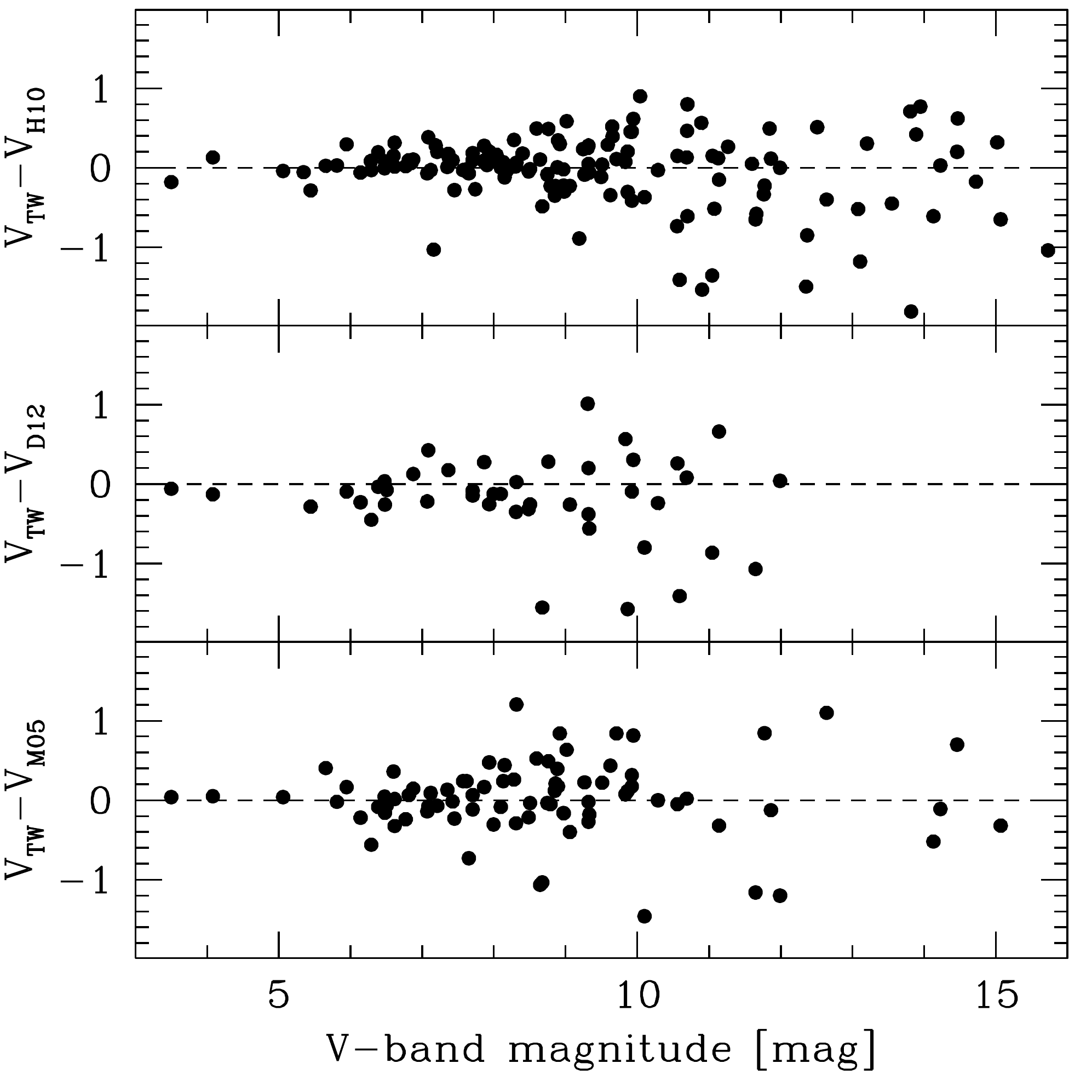}
\caption{Difference in the total magnitudes between this work and (from top to bottom) the 2010 version of \citet{harris1996} (H10), \citet{dalessandroetal2012} (D12) and \citet{mclaughlinvandermarel2005} (M05). The differences increase for fainter
clusters and can reach up to two magnitudes for individual clusters.}
\label{fig:comp1}
\end{center}
\end{figure}

We correct for incompleteness at the faint end by imposing a magnitude cut-off $V_{LIM}$ that is bright enough that the photometry is still complete for stars brighter than $V_{LIM}$ but faint enough that stars 
fainter than $V_{LIM}$ contribute only a small fraction of the cluster light. We usually choose $V_{LIM}$ to be one or two magnitudes below the main-sequence turn-over, depending on the quality of the photometry and the distance of the cluster. This
guarantees that the directly measured bright stars already contribute between 80 to 90\% of the total cluster luminosity. We estimate the contribution of the fainter stars based on the $N$-body models of 
\citet{baumgardthilker2018}.  \citet{baumgardthilker2018} ran a grid of about 3,000 $N$-body simulations and determined
for each cluster the $N$-body model that produced the best fit to the observed stellar mass function at different radii, the observed surface density profile
and the observed velocity dispersion profile. Their simulations provide for each globular cluster a star-by-star model, containing main-sequence, giant stars and compact remnants. For each star we 
use the bolometric luminosity, surface temperature and metallicity from the $N$-body model and convert these into UBVRI magnitudes using the \citet{kurucz1992} atmosphere models
for nuclear burning stars and the bolometric corrections and color indices calculated by \citet{bergeronetal1995} for white dwarfs. After the conversion, we calculate the total V band luminosity,
of all stars $L_{Sim,A}$ and the one for only the bright stars $L_{Sim,B}$ with $V<V_{Lim}$, and correct the observed cluster luminosity by 
$L_{V In} = L_{V Obs} \cdot L_{Sim,A}/L_{Sim,B}$.
Varying $V_{LIM}$ for a few nearby clusters with deep photometry shows that the derived luminosities vary by only about $\pm 0.03$ mag when decreasing the magnitude limit $V_{LIM}$.

In order to estimate the contribution of the cluster parts that are either excluded due to crowding or not covered by the photometry
to the total cluster luminosity, we again use the best-fitting $N$-body model of each cluster and calculate the
total cluster luminosity $L_{Sim,T}$ and the luminosity of the part that is covered by the photometry $L_{Sim,In}$. We then calculate 
the total cluster luminosity by $L_{V, Tot} = L_{V, In} \cdot L_{Sim,T}/L_{Sim,In}$.
Once the total cluster luminosity has been calculated, we calculate the total magnitude of the cluster as $V = -2.5 \cdot \log_{10} L_{V, Tot}$. 
For a few bulge clusters background contamination is an issue even after selecting stars based on CMD position, radial velocities and \gaia proper motions and parallaxes.
In order to correct the cluster luminosity for background stars, we integrate the cluster profile out to large radii so that we can determine the surface
brightness of background stars, and then subtract the total luminosity of the background stars from the cluster luminosity.
Table~\ref{magtab} presents the total cluster luminosities
that we derive this way. It also presents the absolute magnitudes that we calculate from the apparent magnitudes using the
extinction values from \citet{harris1996} as well as the cluster distances from \citet{baumgardtetal2019}, and the radii containing 10\% and 50\%
of the cluster light in projection together with the surface brightness at these radii.

We estimate the error of the cluster luminosity as follows: For clusters where we have no photometry in the Johnson-Cousins system with which to correct the HST magnitudes, we assume an error 
of $\Delta V = 0.10$ mag on the total cluster luminosity. We assume that this error drops to $\Delta V = 0.04$ mag for clusters where we have ground-based V band magnitude measurements. The latter
value is equal to the maximum zero-point uncertainty estimated by \citet{stetsonetal2019} for their photometry.
We also assume that the correction factors $f_{Bright} = L_{Sim,A}/L_{Sim,B} - 1$ and $f_{Field} = L_{Sim,T}/L_{Sim,In}-1$ have 10\% relative errors, 
i.e. if $f_{Bright} = 0.20$ we assume that the corresponding correction of the cluster luminosity has a relative error of 2\%, leading to a 0.022 mag uncertainty of the total cluster magnitude. We also assume that
the global mass function slopes $\alpha$ determined by \citet{baumgardthilker2018} have uncertainties of $\pm$0.20.
Experiments show that the final cluster luminosity changes by 0.02 mag for a change of $\alpha$ of 0.20. Finally, for those clusters where we subtracted a background contribution,
we vary the assumed background level by 10\% and assume an additional magnitude error equal to the change in total cluster magnitude caused by this variation of the assumed
background level.  The total magnitude error is then calculated
combining the various magnitude uncertainties, which we assume to be statistically independent. The magnitude errors are also given in Table~\ref{magtab}.
For the best observed clusters we can achieve errors better than 0.05 mag, i.e. luminosities accurate to about 5\%.

\section{RESULTS }
\label{sec:results}

\subsection{Apparent magnitudes}

Fig.~\ref{fig:comp1} and Table~\ref{tab:comp} compare the apparent magnitudes derived here with those given in the 2010 version of \citet{harris1996}, (H10) \citet{dalessandroetal2012} (D12) and \citet{mclaughlinvandermarel2005} (M05). 
The magnitudes from \citet{harris1996} are mainly derived from aperture photometry, while
\citet{mclaughlinvandermarel2005} derived magnitudes through an integration of the surface density profiles of \citet{trageretal1995}. 
\citet{dalessandroetal2012} derived total magnitudes from data of the GALEX satellite. Hence all estimates use different input data and are more or less independent of each other. We therefore
average the literature estimates and compare them with the magnitudes derived here in the last row of Table~\ref{tab:comp}. Only clusters that have at least two magnitude determinations in the literature are used in the last row.
\begin{table}
\caption{Mean differences and standard deviation around the mean between our photometry and literature values for three different magnitude ranges.} 
\centering
\resizebox{\columnwidth}{!}{%
\begin{tabular}{lcc|cc|cc}
\hline
 Paper & \multicolumn{2}{c|}{$ V< 8$} & \multicolumn{2}{c|}{$ 8 < V < 11$}  & \multicolumn{2}{c}{$ V>11$}   \\
 & $<\!\!\Delta V\!\!>$ &$\sigma_V$ & $<\!\!\Delta V\!\!>$ &$\sigma_V$ &$<\!\!\Delta V\!\!>$ &$\sigma_V$ \\
 & [mag] & [mag] & [mag] & [mag] & [mag] & [mag] \\[+0.05cm]
\hline
Harris (2010)       & $+0.03$ & $0.22$ & $+0.01$ & $0.45$ & $-0.25$ & $0.70$\\[+0.1cm]
Dalessandro et al.  & $-0.07$ & $0.21$ & $-0.25$ & $0.66$ & $-0.31$ & $0.81$\\[+0.1cm]
McLaughlin \& vdM   & $-0.02$ & $0.25$ & $+0.08$ & $0.51$ & $-0.10$ & $0.80$\\[+0.1cm]
Literature averaged & $+0.01$ & $0.15$ & $+0.03$ & $0.35$ & $-0.03$ & $0.53$\\[+0.1cm]
\hline
\end{tabular}}
\label{tab:comp}
\end{table}

The literature values show good agreement with our measurements only for bright clusters with $V<8$ mag, where the average difference is close 
to zero and the typical deviation for individual clusters is about 0.20 mag.
For clusters with total magnitudes fainter than 8 mag, the differences quickly increase and can be as large as 2 magnitudes for some clusters fainter than $V=11$ mag. For all magnitude
ranges, the differences are much larger than
what we expect based on our magnitude errors, indicating that the differences are mainly due to inaccuracies in the literature magnitudes. Interestingly, averaging the literature magnitudes already leads
to significantly smaller scatter than derived from the individual data sets.
A closer examination shows that the most discrepant clusters are often relatively low-mass clusters
like Pal~11 ($M=1.0 \cdot 10^4$ M$_\odot$), Ter~8 ($M=5.9 \cdot 10^4$ M$_\odot$) or Arp~2 ($M=3.9 \cdot 10^4$ M$_\odot$). Since most of these clusters are located in areas of low field-star contamination,
have both deep HST and ground-based photometry, and have proper motions that clearly separate the cluster from the field stars, we regard our total magnitudes as very reliable. The reason for the strong discrepancy 
with the literature values could be that the aperture photometry measurements and the surface brightness profiles of \citet{trageretal1995} have excluded bright giant stars in order to obtain smooth density profiles. 
This could explain why for the faintest clusters our magnitude values are on average smaller (meaning we derive larger total luminosities) compared to the literature values. It could also explain why more 
massive and brighter clusters show better agreement, since the effect of single bright stars is smaller in more massive clusters. The remaining clusters that show large differences
are mostly bulge clusters like NGC~6256, NGC~6453 or NGC~6749 which are located in fields with a large background density of stars. Since we use \gaia proper motions to clean the CMDs from field stars, we again expect our magnitudes to be more reliable than the literature ones.
\begin{figure}[t]
\begin{center}
\includegraphics[width=\columnwidth]{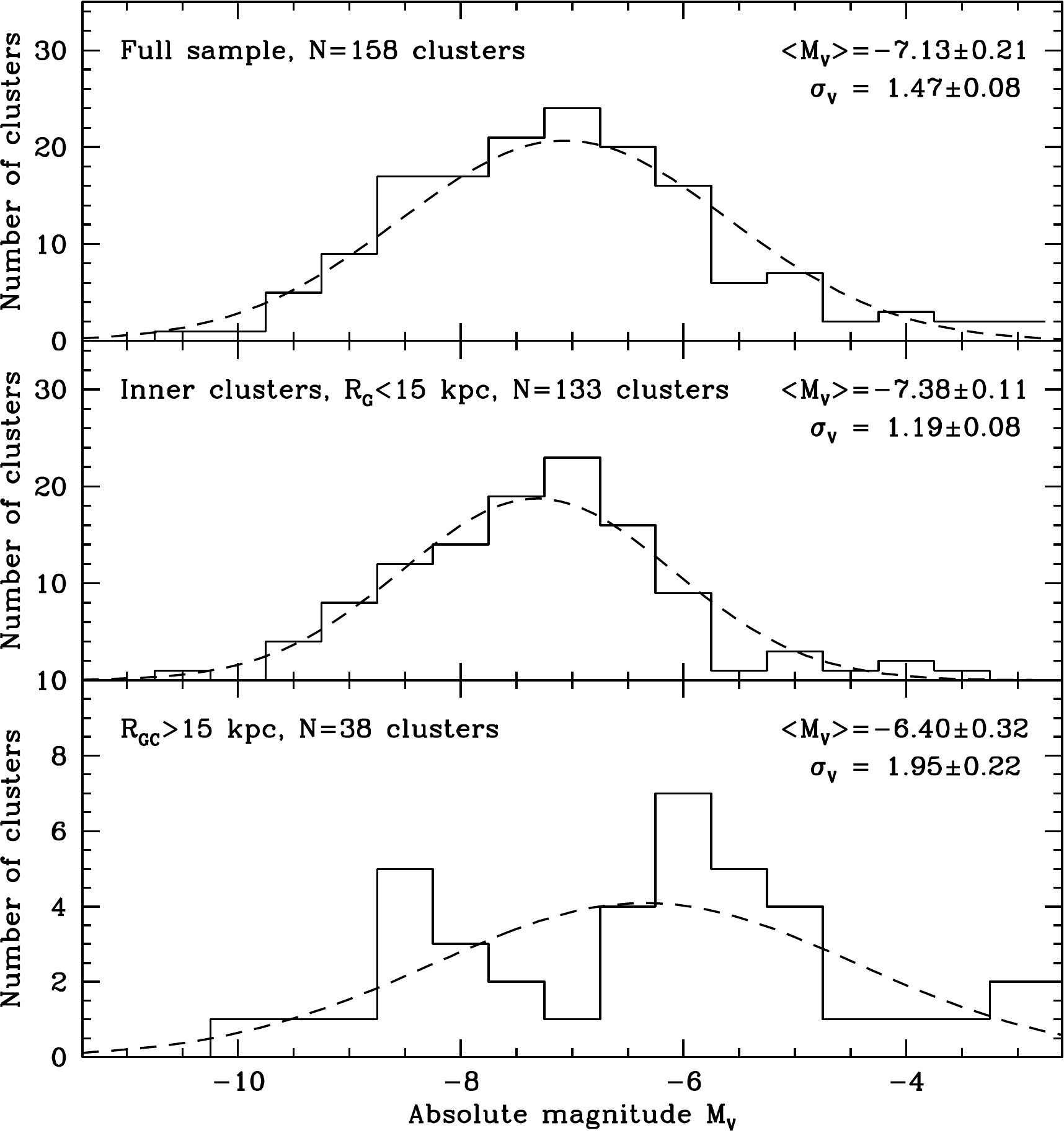}
\caption{Distribution of absolute magnitudes of Milky Way globular clusters. The top panel shows the full distribution, the middle panel the distribution of inner clusters
with $R_G<15$ kpc and the lower panel the distribution of outer clusters with $R_G>15$ kpc. The distribution of outer clusters contains a significantly larger fraction
of low-luminosity clusters.}
\label{fig:mvabs}
\end{center}
\end{figure}

\begin{figure}[t]
\begin{center}
\includegraphics[width=\columnwidth]{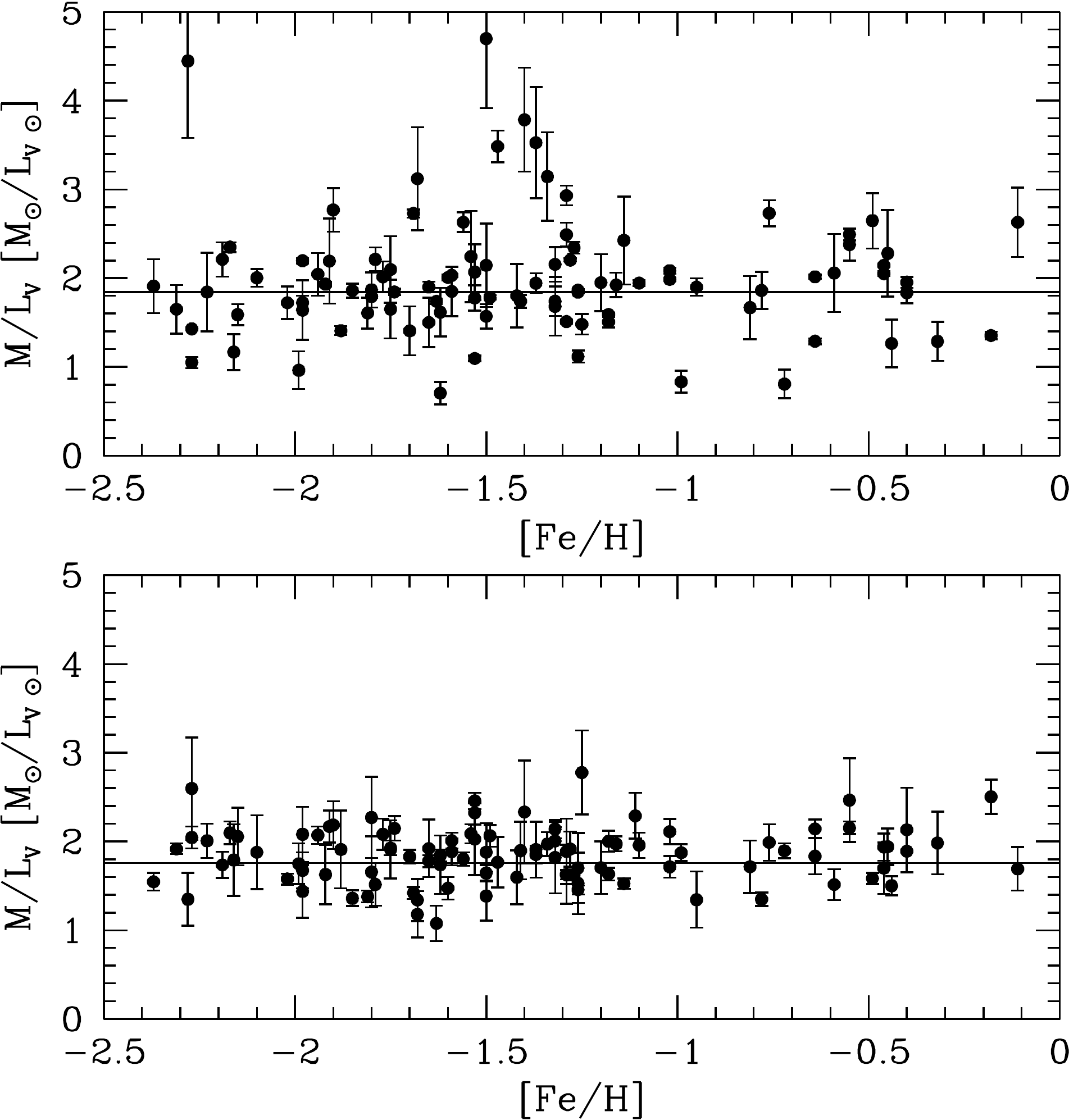}
\caption{$M/L_V$ mass-to-light ratios derived using masses and distances from \citet{baumgardtetal2019}, extinction values from \citet{harris1996} and literature averaged $V$-band luminosities (top panel). The bottom panel shows the mass-to-light ratios derived with the same data but our $V$ band magnitudes. The resulting $M/L_V$ ratios cover a much smaller scatter around the mean value (marked by a solid line).}
\label{fig:mlratio1}
\end{center}
\end{figure}

\begin{figure*}[t]
\begin{center}
\includegraphics[width=0.95\textwidth]{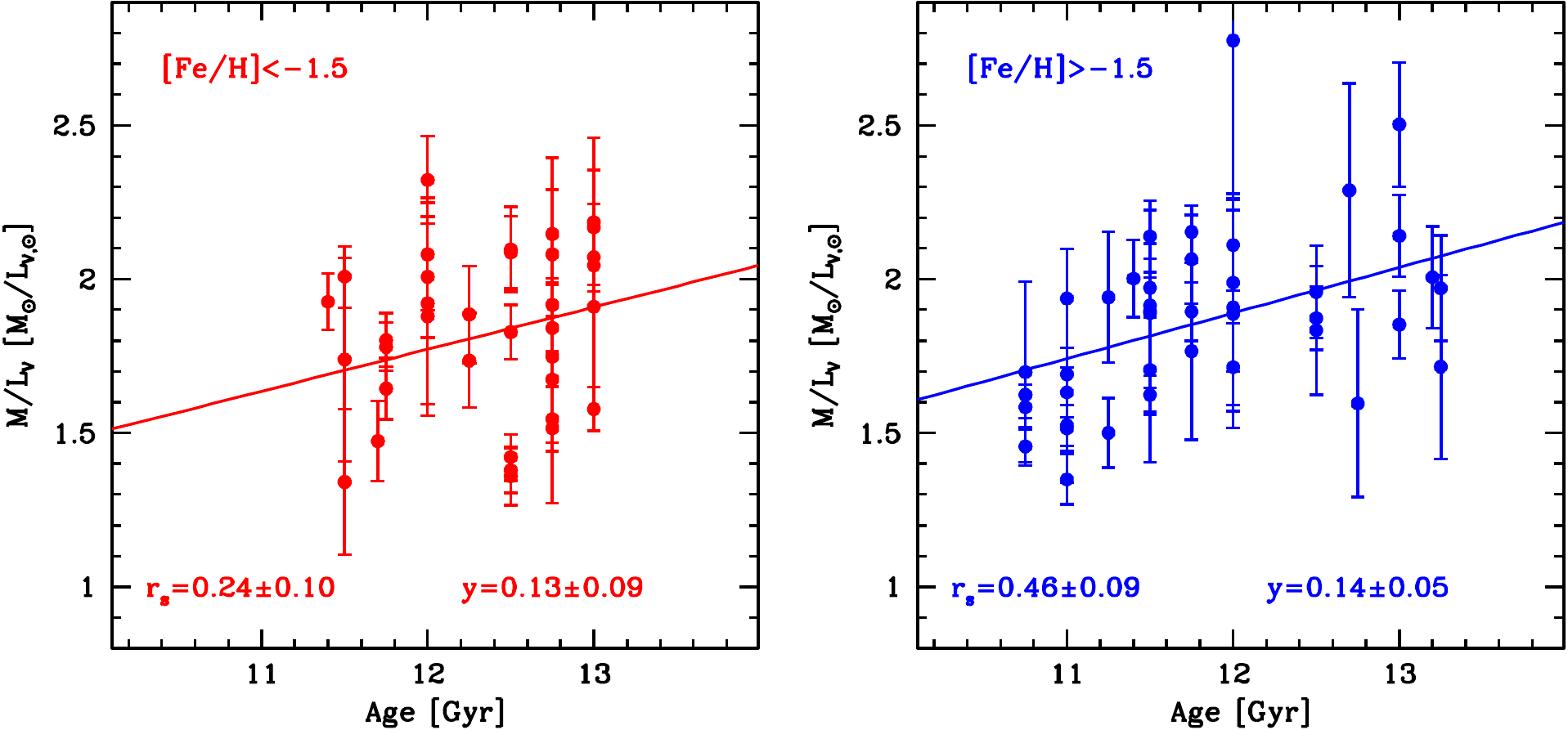}
\caption{Dependence of mass-to-light ratio on the cluster ages for two different metallicity ranges. The $M/L$ ratio increases as a function of age  
in agreement with theoretical predictions. The plots also show the Spearman rank-order co-efficient $r_s$ as well as the slope $y$ of the best-fitting linear 
fit to the data.}
\label{fig:mlratio2}
\end{center}
\end{figure*}

\subsection{Absolute cluster magnitudes}

In order to convert the apparent magnitudes to absolute ones, we use the cluster distances from \citet{baumgardtetal2019} and the $E(B-V)$ values given by \citet{harris1996}. 
Fig.~\ref{fig:mvabs} depicts the distribution of absolute magnitudes of globular clusters that we derive this way, split up into the full sample (upper panel), inner clusters with 
galactocentric distances $R_{GC}<15$ kpc (middle panel) and outer clusters with $R_{GC}>15$ kpc (lower panel). Fitting a Gaussian to the distribution, we obtain a mean magnitude 
of $M_V=-7.13$ and a width $\sigma_V=1.47$ for the full cluster sample.
Our mean absolute magnitude is about 0.2 to 0.3 magnitudes lower and the width is significantly higher than what has been found previously \citep[e.g.][]{secker1992,dicri2006}. The main reason for
the difference are the many low-mass clusters that have been discovered in recent years: For example, out of the 15 globular clusters known today that are not included in the
sample of \citet{dicri2006}, 14 have absolute luminosities below the mean.
It is therefore likely that the average luminosity of MW globular clusters could be lower since our census of globular clusters in the inner parts of the Milky Way is probably
still not complete \citep{minnitietal2017}. The middle and lower panels of
Fig.~\ref{fig:mvabs} split the cluster sample into an inner sample inside $R_{GC}=15$ kpc and an outer one. For the inner clusters we obtain a mean of $M_V=-7.39$ with a width of $\sigma_V=1.19$.
Both values do not seem to depend on the radial range that we consider, indicating that the globular cluster luminosity function is invariant in the inner parts of the Milky Way. In contrast,
the outer clusters have a significantly lower mean luminosity of $M_V=-6.40$ and a much larger width. \citet{huxoretal2014} found evidence that the outer globular cluster system of the Andromeda Galaxy
has a bi-model distribution in luminosity, with a second peak at $M_V-5.5$. The Milky Way globular cluster system could show a similar bi-modality, with one peak at around $M_V=-8$, similar to
the inner globular clusters, and a second peak around $M_V=-6$.   \citet{huxoretal2014} also speculated that the bright outer-halo clusters in M31 formed in-situ, while the clusters in the fainter peak are
accreted from dwarf galaxies. Current orbital data on the Milky Way globular clusters however does not support this conclusion: Out of the 14 globular clusters that are at distances $R_G>$15 kpc and
that are brighter than $M_V=-7.0$, all are associated with a dwarf galaxy progenitor (Gaia-Enceladus, Sagittarius or Helmi Streams) in the recent study by \citet{massarietal2019}. In contrast, among
the 23 outer clusters fainter than $M_V=-7.0$, only 9 are considered securely and 3 are possibly associated with a dwarf galaxy progenitor according to \citet{massarietal2019}.
Hence many of the faint globular clusters could have formed in situ in the halo of the Milky Way without a connection to a dwarf galaxy. Alternatively, they could be associated to dwarf
galaxies that have not yet been discovered.

\subsection{Cluster $M/L_V$ ratios}

Fig.~\ref{fig:mlratio1} compares the mass-to-light ratios that we derive from our magnitudes with the $M/L_V$ ratios that we derive from the literature averaged $V$-band luminosities. In order to
calculate the cluster mass-to-light ratios, we use the cluster masses and distances from \citet{baumgardtetal2019} and the extinction values given by \citet{harris1996}. We only plot clusters that have
relative mass uncertainties $\Delta M/M<0.2$ in Fig.~\ref{fig:mlratio1} in order to be able to better judge the quality of the cluster luminosities. For the same reason, we restrict ourselves to
clusters with extinction values $E(B-V)<1.0$ since for highly reddened clusters differential extinction or uncertainties in the reddening could introduce additional uncertainties. We obtain an average        
mass-to-light ratio of $M/L_V=1.83 \pm 0.03$ $M_\odot/L_\odot$ and a standard deviation of $\sigma_{M/L}=0.24 \pm 0.03$ $M_\odot/L_\odot$ around the mean using our magnitudes compared to 
$M/L_V=1.92 \pm 0.05$ $M_\odot/L_\odot$ and $\sigma_{M/L}=0.49 \pm 0.05$ $M_\odot/L_\odot$ that we derive from the literature magnitudes. Hence, while the average mass-to-light ratio changes by only 
0.1 $M_\odot/L_\odot$, we obtain a much more uniform $M/L_V$ distribution using our magnitudes. In particular, using our magnitudes, no clusters have mass-to-light ratios $M/L_V>2.5$ $M_\odot/L_\odot$ 
or $M/L_V<1$~M$_\odot/L_\odot$, which would be hard to explain using standard isochrones (see sec. 3.2).
\begin{figure*}[t]
\begin{center}
\includegraphics[width=0.95\textwidth]{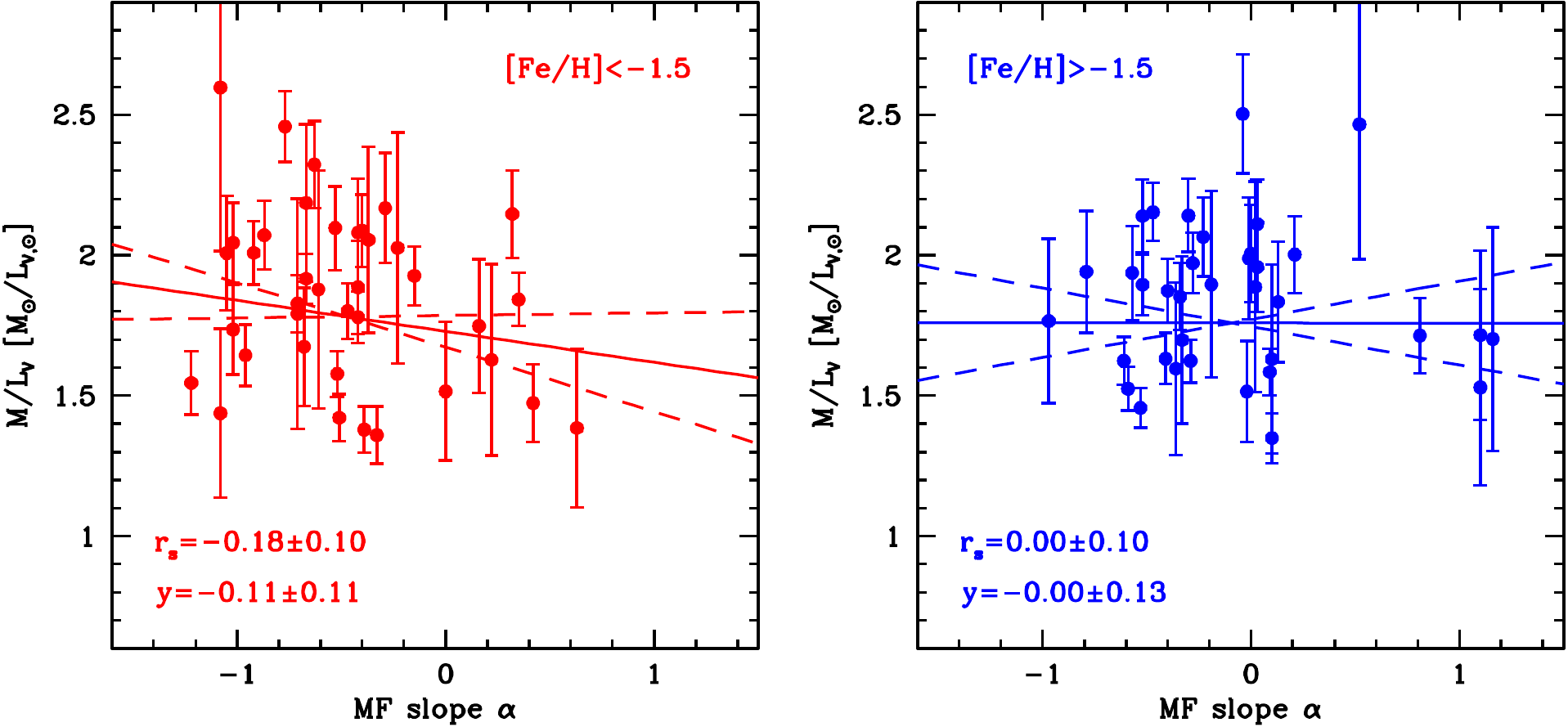}
\caption{Same as Fig.~\ref{fig:mlratio2} but this time showing the dependence of mass-to-light ratio on the mass function of clusters.}
\label{fig:mlratio3}
\end{center}
\end{figure*}

The $M/L$ ratio of a cluster is influenced by several different processes: First, stellar evolution leads to an increase of the $M/L$ ratio with time as massive and bright stars, which are contributing to the total
cluster  light more than to the cluster mass, are constantly being turned into compact 
remnants. According to the PARSEC isochrones \citep{bressanetal2012} stellar evolution should increase the $M/L_V$ ratio of a globular cluster by about 0.3 $M_\odot/L_\odot$ when the cluster
age increases from $T=10$ Gyr to $T=13.5$ Gyr, with only a weak dependence of this increase on metallicity.
Secondly, the stellar mass function of a cluster changes as a result of mass segregation, which lets massive stars sink into the centre and moves low-mass stars towards the outer cluster parts where they
are preferentially removed due to an external tidal field. \citet{baumgardtmakino2003} found by means of $N$-body simulations that this process leads initially to a decrease of the cluster 
$M/L$ ratio since low-mass stars mainly contribute to the cluster mass and only little to the overall cluster light, followed by an increase very close to final dissolution when mainly compact remnants are
left in the cluster.  \citet{baumgardtmakino2003} found a maximum decrease of the $M/L$ ratio of about $\Delta M/L_V$=0.5 to 0.7 due to mass loss. A decrease of the cluster $M/L$ ratio during the evolution was also found by \citet{bianchinietal2017},
although they found that the ejection of dark remnants is also important in changing the $M/L$ ratio of a star cluster. Finally, $M/L$ ratios also depend on the metallicity of a cluster.
\citet{maraston1999} predicted an increase of the V-band $M/L$ ratio by a factor three when increasing the metallicity from [Fe/H]=-0.5 to [Fe/H]=+0.3.

Figs.~\ref{fig:mlratio2} and \ref{fig:mlratio3} depict the dependence of the $M/L$ ratios that we derive from our magnitudes on the cluster age and mass function slope $\alpha$. Similar to the previous plots, we
show only clusters that have relative mass errors $\Delta M/M<0.2$ and reddening values $E(B-V)<1.0$. We also divide the sample into low-metallicity clusters with [Fe/H]<-1.5 and high-metallicity clusters
with [Fe/H]>-1.5 to reduce the dependency on metallicity. We have taken the cluster ages in Fig.~\ref{fig:mlratio2} mainly from \citet{vandenbergetal2013}, or, if a cluster was not studied by them, from the literature.
We obtain highly significant positive Spearman rank-order coefficients $r_s$ for both metallicity ranges. Also a linear fit of the form $M/L_V = x+y \cdot T_{Age}$
gives positive slopes $y$ for both metallicity ranges, indicating that cluster $M/L$ ratios increase with age. The increase seen when going from T=10 Gyr to T=13.5 Gyr is about 0.45 $M_\odot/L_\odot$ for 
both metal-poor and metal-rich clusters, in agreement with the predicted change based on stellar isochrones.

Fig.~\ref{fig:mlratio3} depicts the $M/L_V$ ratios against the mass function of the clusters, taken from \citet{baumgardtetal2019}. We obtain a weak anti-correlation between the mass-to-light ratio of a 
cluster and its mass function slope $\alpha$
only for the metal-poor clusters. For the metal-rich clusters there is no visible correlation. One reason for the lack of a correlation for metal-rich clusters could be that either the mass or mass function 
measurements for these have large errors since many of these clusters are located in the bulge, which makes observations of them more difficult. Alternatively, clusters could start with different 
initial mass functions, so the present-day differences in the MF slope $\alpha$ are not due to dynamical evolution.

\subsection{Contribution of different stars to the total magnitudes}

\begin{figure*}[h]
\begin{center}
\includegraphics[width=0.95\textwidth]{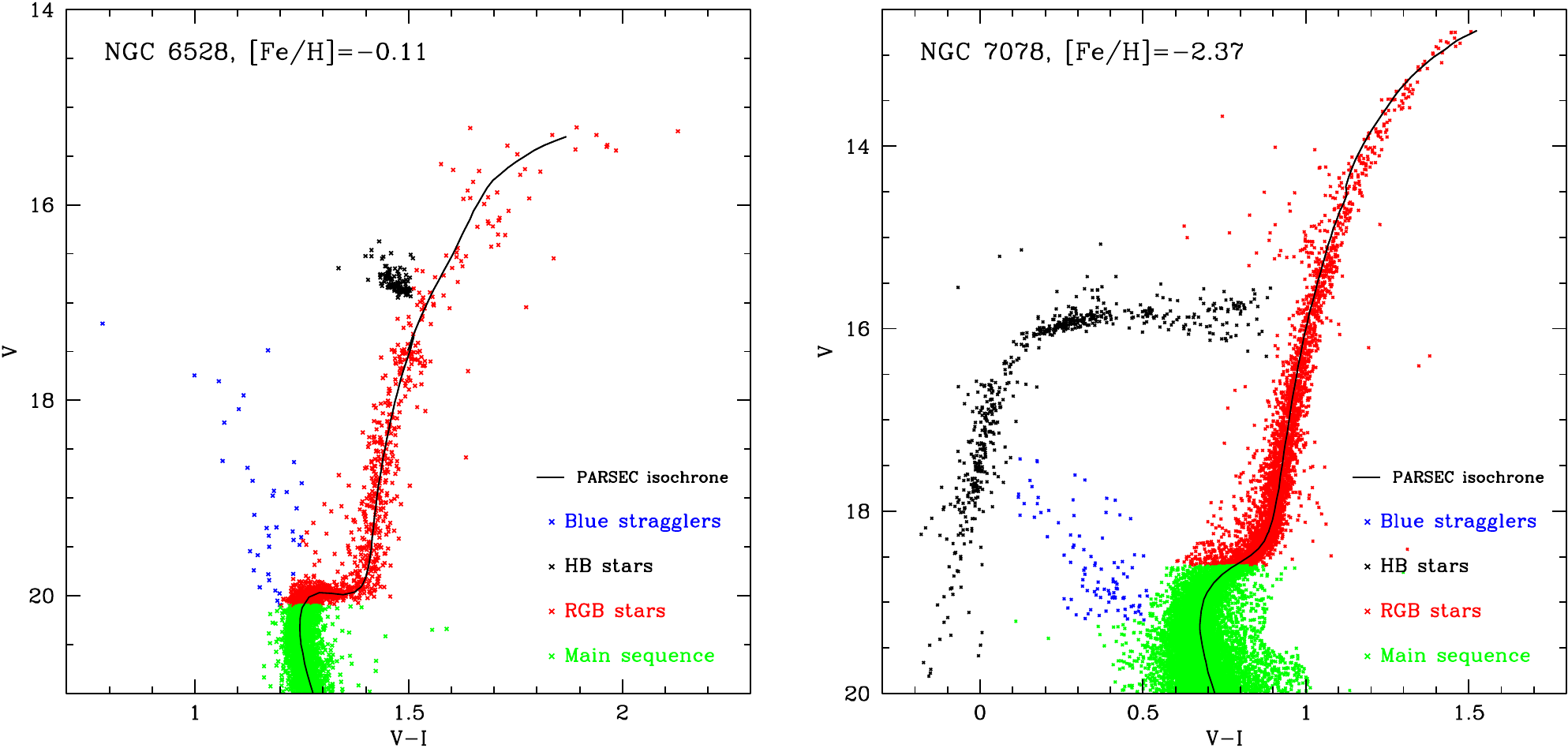}
\caption{Illustration of our division of stars into different evolutionary stages for the high metallicity cluster NGC~6528 (left panel) and the low metallicity cluster NGC~7078 (right panel). Stars are split
into blue stragglers (blue), horizontal branch stars (HB, black), red giant branch stars (RGB, red) and main sequence stars (green). Only stars that pass the various membership criteria detailed in sec.~2 are shown.}
\label{fig:evo}
\end{center}
\end{figure*}

Table~\ref{tab:magsplit} shows the contribution of stars in different evolutionary stages to the total $V$-band magnitudes. We have split stars into main sequence stars (MS),
horizontal branch stars (HB), blue stragglers (BS) and red giant branch stars (RGB), where the RGB stars include asymptotic giant branch and sub-giant branch stars as well.
As an example, Fig.~\ref{fig:evo} depicts the division of stars for two clusters, NGC~6528 and NGC~7078.
We have analyzed six clusters, roughly equally spaced in metallicty between $[Fe/H]=-2.37$ to $[Fe/H]=-0.11$, thereby encompassing the range of metallicities seen
for Galactic globular clusters. 
We again use our $N$-body models to correct for main-sequence stars too faint to be seen in the observations.

It can be seen that blue stragglers contribute only about $\sim$1\% to the total cluster light, with a strong increase of their contribution towards higher metallicity. The
fraction of light from HB stars is also increasing with metallicity, while the fraction of light coming from the RGB is roughly constant at around 55\%. The increase of 
the fraction of light in HB stars
at increasing metallicity confirms predictions by stellar evolution models \citep[e.g.][]{renzinibuzzoni1986}, which predict such an increase as a result
of the shorter evolutionary timescales on the RGB stars. There also seems to be a slight decrease of the fraction of light from MS stars with metallicity, however this decrease could
also be driven by other factors like a change of the internal mass function of the clusters. A larger sample would be needed to disentangle the different effects.
\begin{table}
\caption{Relative contribution of stars in different evolutionary stages to the total $V$-band magnitudes for six clusters.}
\centering
\resizebox{\columnwidth}{!}{%
\begin{tabular}{lcccccc}
\hline
\multirow{2}{*}{Cluster} & \multirow{2}{*}{[Fe/H]} & BS  & HB & RGB & MS \\
 & & [\%] & [\%] & [\%] & [\%] \\ 
\hline
NGC 6528 &  -0.11  &  1.3  &  22.4  & 51.1   & 25.0 \\
NGC 6496 &  -0.46  &  0.8  &  12.2  & 59.8   & 27.2 \\
NGC 6171 &  -1.02  &  0.9  &  13.8  & 57.1   & 28.3 \\
NGC 5272 &  -1.50  &  0.3  &  10.9  & 55.9   & 32.9 \\
NGC 6397 &  -2.02  &  0.3  &  $\,\,\,$9.5  & 60.8   & 29.3 \\
NGC 7078 &  -2.37  &  0.2  &  $\,\,\,$8.9  & 52.4   & 38.6 \\
\hline
\end{tabular}}
\label{tab:magsplit}
\end{table}

\subsection{Comparison with stellar evolution models}

\begin{figure*}[t]
\begin{center}
\includegraphics[width=\textwidth]{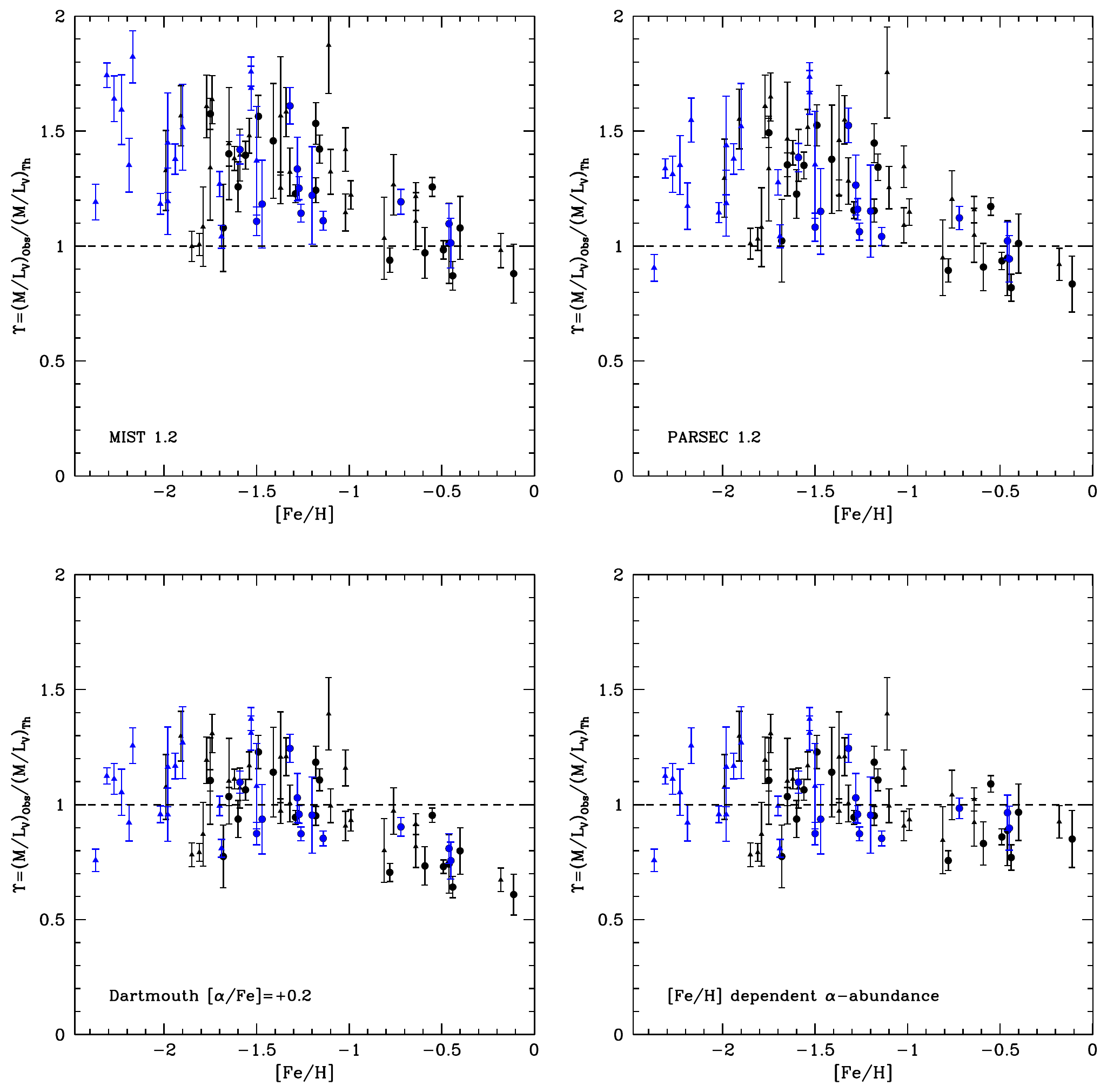}
\caption{Ratio of the measured $M/L_V$ ratios to the $M/L_V$ ratio predicted by different stellar-evolution models as a function of the cluster metallicity. Shown is a comparison against MIST isochrones (top left), PARSEC isochrones (top right), Dartmouth isochrones with enhanced $\alpha$ element ratios of [$\alpha$/Fe]=+0.2 and a model in which the $\alpha$-element abundances decrease for
metal rich clusters (lower right).}
\label{fig:mlratio4}
\end{center}
\end{figure*}

Fig.~\ref{fig:mlratio4} depicts the ratio of the $M/L_V$ values that we derive from our apparent cluster magnitudes with the predictions of stellar evolution models. Shown are predictions
from version 1.2 of the MIST isochrones \citep{paxtonetal2015, dotter2016, choietal2016}, version 1.2 of the PARSEC isochrones \citep{bressanetal2012} and predictions from the
DARTMOUTH isochrones \citep{dotteretal2008}. For each stellar evolution model, we created a two-dimensional grid of isochrones in age and metallicity. Isochrones were spaced by 0.5 Gyr 
between 10 Gyr and 13.5 Gyr in age and by $\Delta [Fe/H]=0.10$ between $[Fe/H]=-2.50$ and $[Fe/H]=0.0$ in metallicity. For each cluster age, metallicity and stellar evolution model, we then 
distributed $5 \cdot 10^5$ stars following the mass functions given in \citet{baumgardthilker2018} and then used the isochrones to calculate the total luminosity, total mass and $M/L$ ratio 
of each model. We thus obtained a grid of 1,456 models giving the $M/L_V$ ratio of star clusters as a function of cluster age, metallicity and internal mass function for each stellar evolution 
model. We then linearly interpolate in this grid of models to predict the expected mass-to-light ratio of each globular cluster given its mass function, metallicity and cluster age. 

Fig.~\ref{fig:mlratio4}  compares the ratio of the observed $M/L_V$ ratios to the theoretically predicted ones. It can be seen that the predictions of the MIST and PARSEC isochrones are 
very similar. Both reproduce the observed mass-to-light ratios of metal-rich clusters quite well, but predict higher than observed $M/L_V$ ratios for the metal-poor clusters.
The mismatch between observed and predicted $M/L_V$ ratios of metal-rich globular clusters noted by \citet{straderetal2011} is therefore mainly due to the fact that \citet{straderetal2011} assumed a
Kroupa or Salpeter type mass function for the clusters, while the data for the Milky Way globular clusters indicates that they follow much shallower mass functions. 
At the moment, the data does not support a metallicity dependent top-heavy IMF as suggested by \citet{zonoozietal2016}, although we can't rule out any such variation either.

A possible reason for the mismatch between predicted and observed $M/L_V$ ratios of metal-poor GCs could be that the isochrone models considered a solar-abundance pattern while metal poor
stars are known to be enriched in $\alpha$-elements \citep{pritzletal2005}. The lower-left hand figure therefore compares our observed $M/L_V$ ratios with DARTMOUTH isochrones that
have an $\alpha$-element enhancement of $[\alpha/Fe]=+0.2$. We use DARTMOUTH isochrones since for MIST and PARSEC isochrones only solar abundance models are available. It can be seen that
for $\alpha$-enhanced isochrones the predicted $M/L_V$ ratios of metal-poor clusters with $[Fe/H]<-1$ are in agreement with the observed ones, while the
metal-rich clusters now have too low $M/L_V$ ratios. However there are indications that the $\alpha$-element enhancement of globular clusters decreases for clusters with $[Fe/H]>-1$ down
to solar values \citep[e.g.][]{ferraroetal1999,pritzletal2005}. Fig.~3 in \citet{hortaetal2020} for example shows that clusters with  $[Fe/H]<-0.7$ have more or less constant $[Si/Fe]$ of about $[Si/Fe]=+0.25$, followed by a downturn
in the $[Si/Fe]$ values similar to what is seen for field stars. The two most metal-rich clusters in their sample (Liller~1 and Pal~10) have $[Si/Fe]$ of $0.01 \pm 0.05$ and
$0.0 \pm 0.10$ respectively, i.e. they are compatible with a solar abundance ratio. A downturn of the $\alpha$-element abundances for metal-rich GCs is also predicted based on
cosmological simulations \citep[e.g.][]{hughesetal2020}.

We therefore adopt an $\alpha$-element distribution with $[\alpha/Fe]=+0.2$ for clusters with
$[Fe/H]<-0.8$ followed by a linear decrease down to $[\alpha/Fe]=+0.0$ for $[Fe/H]=0.0$ and interpolate in our grid between the $[\alpha/Fe]=+0.2$ DARTMOUTH models and the
PARSEC models depending on the $\alpha$-element enhancement of each cluster. The resulting $M/L_V$ ratios are shown in the lower right corner. It can be seen that we now obtain  
an excellent agreement between the observed and expected mass-to-light ratios at all metallicities. 

\section{CONCLUSION}
\label{sec:concl}

We have derived new $V$-band magnitudes of 153 Galactic globular clusters by summing up the magnitudes of their individual members stars derived from HST and ground-based photometry
and correcting the derived magnitudes for missing faint stars and spatial incompleteness. In order to differentiate
between cluster and field stars, we have made use of the positions of stars in colour-magnitude diagrams, available radial-velocity information and proper motions and parallaxes from \gaia DR2. Our $V$-band magnitudes 
show good agreement with published literature magnitudes 
for bright clusters with $V<8$ mag. For fainter clusters, typical differences are of order 0.5 mag with individual clusters showing differences of up to 2 magnitudes.

Our $V$-band magnitudes lead to a mean mass-to-light ratio of $M/L_V=1.83$ and a scatter of $\sigma_V=0.24$ $M_\odot/L_\odot$ around the mean,
significantly smaller then the scatter obtained with literature magnitudes. In agreement with \citet{straderetal2011}, we find no dependence of the average mass-to-light ratio of a cluster with metallicity.
We find evidence that the mass-to-light ratios of globular clusters are increasing with cluster age, in agreement with theoretical predictions. 
We also find good agreement between the derived mass-to-light ratios with the expected ones from stellar isochrones if the mass function of the clusters is taken into account. PARSEC and MIST stellar isochrones 
with a solar-abundance pattern for $\alpha$-elements are able to reproduce the mass-to-light ratios only for clusters with $[Fe/H]>-1.0$, but predict too high mass-to-light ratios for the more metal-poor clusters. However
using $\alpha$-enhanced DARTMOUTH isochrones with $[\alpha/Fe]=+0.2$ leads to a good agreement of observed and predicted $M/L$ ratios. There is therefore no evidence for a significant amount of dark matter
inside the main body of globular clusters as has been previously suggested \citep{baumgardtmieske2008,wirth2020}. However the data does not rule out dark matter haloes surrounding globular clusters or 
a small amount (of order 20\% of the total mass or less) of dark matter inside the clusters.

We finally find that globular clusters at galactocentric distances $R_G>15$ kpc have on average about one magnitude lower absolute magnitudes than clusters inside this radius. This could be either due to
the weaker tidal field in the outer parts of the Milky Way, which increases the lifetime of low mass clusters to more than a Hubble time, or due to the fact that low-mass clusters in the inner Milky Way have not yet been found due to 
large reddening and the strong background density of stars.
About half of the low-luminosity clusters in the outer parts are not connected to any known dwarf galaxy or any known past merger events that are thought to have happened in the early Milky Way, indicating that
they either formed in-situ or are connected to so far undiscovered dwarf galaxies.

In the future, determining total magnitudes, cluster colours and $M/L$ ratios in other wavelength bands will be useful for tests of stellar evolution and galactic studies. 
Given our photometric data and the fact that we already have determined the cluster members, this task should also be straightforward to implement. 
Our photometry should also help determine the contribution of stars in different evolutionary stages (RGB, HB, main sequence and blue stragglers) to the total magnitudes in the different bands.
The caveat, however, is that individual stellar photometry in bands other than the V band is currently only 
available for a subset of Galactic clusters.

\begin{acknowledgements}
We thank Sylvie Beaulieu, Aaron Dotter, Matthias Frank, Edoardo P. Lagioia, Domenico Nardiello, Stefano Oliveira de Souza, Soeren S. Larsen, Sebastian Kamann, Jeffrey Simpson and Daniel Weisz for sharing their globular cluster photometry with us. We also acknowledge
the work by Peter Stetson in deriving wide-field, ground-based photometry of globular clusters and making his data public before publication. We finally thank Mirek Giersz and Jarrod Hurley for 
making a FORTRAN code available to us that converts bolometric luminosities and temperatures into absolute magnitudes in the Johnson-Cousins system. Based on observations made with the NASA/ESA Hubble Space Telescope, 
obtained from the data archive at the Space Telescope Science Institute. STScI is operated by the Association of Universities for Research in Astronomy, Inc. under NASA contract NAS 5-26555. This research has made use of the SIMBAD database, operated at CDS, Strasbourg, France. This work has made use of data from the European Space Agency (ESA) mission
{\it Gaia} (\url{https://www.cosmos.esa.int/gaia}), processed by the {\it Gaia}
Data Processing and Analysis Consortium (DPAC,
\url{https://www.cosmos.esa.int/web/gaia/dpac/consortium}). Funding for the DPAC
has been provided by national institutions, in particular the institutions
participating in the {\it Gaia} Multilateral Agreement.
\end{acknowledgements}

\begin{appendix}

\section{Input photometry used to derive the apparent magnitudes of globular clusters}

\begin{table}
\caption{Input photometry used to calculate the total magnitudes of GCs} 
\centering
\resizebox{\columnwidth}{!}{%
\begin{tabular}{@{}lll@{}}
\hline\hline
Cluster & Source of Photometry & Telescope/Instrument/Band \\
\hline\hline
AM 1     & \citet{dotteretal2008b} & HST/WFPC2 F555W/F814W \\
         & \citet{hilker2006} & ESO VLT/FORS2 BV\\[+0.05cm]
AM 4     & This work & HST/WFC3 F606W/F814W \\
         & \citet{inmancarney1987} & CTIO BV \\[+0.05cm]
Arp 2    & \citet{sarajedinietal2007} & HST/ACS F606W/F814W \\
         & \citet{stetson2020} & Ground-based UBVRI \\[+0.05cm]
BH 261   & \citet{carraroetal2005} & Ground-based BVI \\
         & \citet{schlaflyetal2018} & Ground-based ugrizY \\[+0.05cm]
Crater   & \citet{weiszetal2016} & HST/ACS F606W/F814W \\[+0.05cm]
Djorg 1  & \citet{ortolanietal2019} & HST/ACS/WFC3 F606W/F160W \\
         & \citet{ortolanietal1995} & ESO NTT VI \\[+0.05cm]
Djorg 2  & \citet{ortolanietal2019b}  & HST/ACS/WFC3 F606W/F110W \\
         & \citet{ortolanietal1997c} & Ground-based VI \\[+0.05cm]
E 3      & \citet{sarajedinietal2007} & HST/ACS F606W/F814W \\
         & \citet{stetsonetal2019} & Ground-based UBVRI \\[+0.05cm]
Eridanus & \citet{stetsonetal1999} & HST/WFPC2 F555W/F814W \\
         & \citet{stetson2020} & Ground-based UBVRI \\
         & \citet{munozetal2018c} & CFHT/MegaCam g/r \\[+0.05cm]
ESO 280-SC06 & \citet{simpson2018} & Ground-based VI \\[+0.05cm]
%%ESO 452-SC11 & \citet{chambersetal2016} & Ground-based grizy \\[+0.05cm]
ESO 452-SC11 & This work & Ground-based VI \\[+0.05cm]
FSR 1735 & This work & HST/ACS/WFC3 F606W/F110W \\[+0.05cm]
FSR 1758 & \citet{schlaflyetal2018} & Ground-based ugrizY \\[+0.05cm]
HP 1     & \citet{schlaflyetal2018} & Ground-based ugrizY \\
         & \citet{stetson2020} & Ground-based UBVRI \\[+0.05cm]
IC 1257  & This work & Ground-based VI \\[+0.05cm]
IC 1276  & \citet{stetson2020} & Ground-based UBVRI \\[+0.05cm]
IC 4499  & \citet{dotteretal2011} & HST/ACS F606W/F814W \\
         & \citet{stetsonetal2019} & Ground-based UBVRI \\[+0.05cm]
Lynga 7  & \citet{sarajedinietal2007} & HST/ACS F606W/F814W \\[+0.05cm]
NGC 104  & \citet{sarajedinietal2007} & HST/ACS F606W/F814W \\
         & \citet{stetsonetal2019} & Ground-based UBVRI \\[+0.05cm]
NGC 288  & \citet{sarajedinietal2007} & HST/ACS F606W/F814W \\
         & \citet{stetsonetal2019} & Ground-based UBVRI \\[+0.05cm]
NGC 362  & \citet{sarajedinietal2007} & HST/ACS F606W/F814W \\
         & \citet{stetson2020} & Ground-based UBVRI \\[+0.05cm]
NGC 1261 & \citet{sarajedinietal2007} & HST/ACS F606W/F814W \\
         & \citet{stetsonetal2019} & Ground-based UBVRI \\[+0.05cm]
NGC 1851 & \citet{sarajedinietal2007} & HST/ACS F606W/F814W \\
         & \citet{stetsonetal2019} & Ground-based UBVRI \\[+0.05cm]
NGC 1904 & \citet{piottoetal2002} & HST/WFPC2 F439W/F555W \\
         & \citet{stetsonetal2019} & Ground-based UBVRI \\[+0.05cm]
NGC 2298 & \citet{sarajedinietal2007} & HST/ACS F606W/F814W \\
         & \citet{stetsonetal2019} & Ground-based UBVRI \\[+0.05cm]
NGC 2419 & \citet{larsenetal2019} & HST/ACS F438W/F555W \\
         & \citet{beccarietal2013} & LBT, uVI \\
         & \citet{stetson2020} & Ground-based UBVRI \\[+0.05cm]
NGC 2808 & \citet{sarajedinietal2007} & HST/ACS F606W/F814W \\
         & \citet{stetsonetal2019} & Ground-based UBVRI \\[+0.05cm]
NGC 3201 & \citet{sarajedinietal2007} & HST/ACS F606W/F814W \\
         & \citet{stetsonetal2019} & Ground-based UBVRI \\[+0.05cm]
NGC 4147 & \citet{sarajedinietal2007} & HST/ACS F606W/F814W \\
         & \citet{stetsonetal2019} & Ground-based UBVRI \\[+0.05cm]
NGC 4372 & \citet{piottoetal2002} & HST/WFPC2 F439W/F555W \\
         & \citet{stetsonetal2019} & Ground-based UBVRI \\[+0.05cm]
NGC 4590 & \citet{sarajedinietal2007} & HST/ACS F606W/F814W \\
         & \citet{stetsonetal2019} & Ground-based UBVRI \\[+0.05cm]
NGC 4833 & \citet{sarajedinietal2007} & HST/ACS F606W/F814W \\
         & \citet{stetsonetal2019} & Ground-based UBVRI \\[+0.05cm]
NGC 5024 & \citet{sarajedinietal2007} & HST/ACS F606W/F814W \\
         & \citet{stetsonetal2019} & Ground-based UBVRI \\[+0.05cm]
NGC 5053 & \citet{sarajedinietal2007} & HST/ACS F606W/F814W \\
         & \citet{stetsonetal2019} & Ground-based UBVRI \\[+0.05cm]
NGC 5139 & \citet{sarajedinietal2007} & HST/ACS F606W/F814W \\
         & \citet{stetsonetal2019} & Ground-based UBVRI \\[+0.05cm]
NGC 5272 & \citet{sarajedinietal2007} & HST/ACS F606W/F814W \\
         & \citet{stetsonetal2019} & Ground-based UBVRI \\[+0.05cm]
NGC 5286 & \citet{sarajedinietal2007} & HST/ACS F606W/F814W \\
         & \citet{stetsonetal2019} & Ground-based UBVRI \\[+0.05cm]
NGC 5466 & \citet{sarajedinietal2007} & HST/ACS F606W/F814W \\
         & \citet{stetson2020} & Ground-based UBVRI \\[+0.05cm]
\hline\hline
\end{tabular}}
\label{tab1}
\end{table}

\addtocounter{table}{-1}

\begin{table}
\caption{continued}
\centering
\resizebox{\columnwidth}{!}{%
\begin{tabular}{@{}lll@{}}
\hline\hline
Cluster & Source of Photometry & Telescope/Instrument/Band \\
\hline\hline
NGC 5634 & \citet{piottoetal2002} & HST/WFPC2 F439W/F555W \\
         & \citet{stetsonetal2019} & Ground-based UBVRI \\[+0.05cm]
NGC 5694 & \citet{piottoetal2002} & HST/WFPC2 F439W/F555W \\
         & \citet{stetsonetal2019} & Ground-based UBVRI \\[+0.05cm]
NGC 5824 & \citet{piottoetal2002} & HST/WFPC2 F439W/F555W \\
         & \citet{stetsonetal2019} & Ground-based UBVRI \\[+0.05cm]
NGC 5897 & \citet{nardielloetal2018} & HST/WFC3 F606W/F814W \\
         & \citet{stetson2020} & Ground-based UBVRI \\[+0.05cm]
NGC 5904 & \citet{sarajedinietal2007} & HST/ACS F606W/F814W \\
         & \citet{stetsonetal2019} & Ground-based UBVRI \\[+0.05cm]
NGC 5927 & \citet{sarajedinietal2007} & HST/ACS F606W/F814W \\
         & \citet{stetsonetal2019} & Ground-based UBVRI \\[+0.05cm]
NGC 5946 & \citet{baumgardtetal2019} & HST/WFC3 F438W/F555W \\
         & \citet{alcainoetal1991} & Ground-based BV \\[+0.05cm]
NGC 5986 & \citet{sarajedinietal2007} & HST/ACS F606W/F814W \\
         & \citet{stetsonetal2019} & Ground-based UBVRI \\[+0.05cm]
NGC 6093 & \citet{sarajedinietal2007} & HST/ACS F606W/F814W \\[+0.05cm]
         & \citet{stetson2020} & Ground-based UBVRI \\[+0.05cm]
NGC 6101 & \citet{sarajedinietal2007} & HST/ACS F606W/F814W \\
         & \citet{stetsonetal2019} & Ground-based UBVRI \\[+0.05cm]
NGC 6121 & \citet{sarajedinietal2007} & HST/ACS F606W/F814W \\
         & \citet{stetsonetal2019} & Ground-based UBVRI \\[+0.05cm]
NGC 6139 & \citet{baumgardtetal2019} & HST/WFC3 F438W/F555W \\
         & \citet{stetson2020} & Ground-based UBVRI \\[+0.05cm]
NGC 6144 & \citet{sarajedinietal2007} & HST/ACS F606W/F814W \\
         & \citet{stetson2020} & Ground-based UBVRI \\[+0.05cm]
NGC 6171 & \citet{sarajedinietal2007} & HST/ACS F606W/F814W \\
         & \citet{stetson2020} & Ground-based UBVRI \\[+0.05cm]
NGC 6205 & \citet{sarajedinietal2007} & HST/ACS F606W/F814W \\
         & \citet{stetsonetal2019} & Ground-based UBVRI \\[+0.05cm]
NGC 6218 & \citet{sarajedinietal2007} & HST/ACS F606W/F814W \\
         & \citet{stetsonetal2019} & Ground-based UBVRI \\[+0.05cm]
NGC 6229 & \citet{piottoetal2002} & HST/WFPC2 F439W/F555W \\
         & \citet{stetson2020} & Ground-based UBVRI \\[+0.05cm]
NGC 6235 & \citet{piottoetal2002} & HST/WFPC2 F439W/F555W \\
         & \citet{stetson2020} & Ground-based UBVRI \\[+0.05cm]
NGC 6254 & \citet{sarajedinietal2007} & HST/ACS F606W/F814W \\
         & \citet{stetsonetal2019} & Ground-based UBVRI \\[+0.05cm]
NGC 6256 & \citet{baumgardtetal2019} & HST/WFC3 F555W/F814W \\
         & \citet{stetson2020} & Ground-based UBVRI \\[+0.05cm]
NGC 6266 & \citet{piottoetal2002} & HST/WFPC2 F439W/F555W \\
         & \citet{stetson2020} & Ground-based UBVRI \\[+0.05cm]
NGC 6273 & \citet{baumgardtetal2019} & HST/WFC3 F555W/F814W \\
         & \citet{stetson2020} & Ground-based UBVRI \\[+0.05cm]
NGC 6284 & \citet{piottoetal2002} & HST/WFPC2 F439W/F555W \\
         & \citet{stetson2020} & Ground-based UBVRI \\[+0.05cm]
NGC 6287 & \citet{piottoetal2002} & HST/WFPC2 F439W/F555W \\
         & \citet{stetson2020} & Ground-based UBVRI \\[+0.05cm]
NGC 6293 & \citet{kamannetal2018} & HST/WFC3 F555W/F814W \\[+0.05cm]
NGC 6304 & \citet{sarajedinietal2007} & HST/ACS F606W/F814W \\
         & \citet{rosenbergetal2000} & ESO Dutch Telescope VI \\[+0.05cm]
NGC 6316 & \citet{piottoetal2002} & HST/WFPC2 F439W/F555W \\
         & \citet{laydenetal2003}  & CTIO VI \\[+0.05cm]
NGC 6325 & \citet{baumgardtetal2019} & HST/WFC3 F438W/F555W \\[+0.05cm]
         & \citet{stetson2020} & Ground-based UBVRI \\[+0.05cm]
NGC 6333 & \citet{baumgardtetal2019} & HST/ACS F435W/F555W \\
         & \citet{stetson2020} & Ground-based UBVRI \\[+0.05cm] 
NGC 6341 & \citet{sarajedinietal2007} & HST/ACS F606W/F814W \\
         & \citet{stetsonetal2019} & Ground-based UBVRI \\[+0.05cm]
NGC 6342 & \citet{baumgardtetal2019} & HST/WFC3 F438W/F555W \\[+0.05cm] 
         & \citet{stetsonetal2019} & Ground-based UBVRI \\[+0.05cm]
NGC 6352 & \citet{sarajedinietal2007} & HST/ACS F606W/F814W \\[+0.05cm]
         & \citet{stetson2020} & Ground-based UBVRI \\[+0.05cm]
\hline\hline
\end{tabular}}
\end{table}

\addtocounter{table}{-1}

\begin{table}
\caption{continued}
\centering
\resizebox{\columnwidth}{!}{%
\begin{tabular}{@{}lll@{}}
\hline\hline
Cluster & Source of Photometry & Telescope/Instrument/Band \\
\hline\hline
NGC 6355 & \citet{baumgardtetal2019} & HST/WFC3 F438W/F555W \\[+0.05cm]
NGC 6356 & \citet{piottoetal2002} & HST/WFPC2 F439W/F555W \\[+0.05cm]
         & \citet{stetson2020} & Ground-based UBVRI \\[+0.05cm]
NGC 6362 & \citet{sarajedinietal2007} & HST/ACS F606W/F814W \\
         & \citet{stetson2020} & Ground-based UBVRI \\[+0.05cm]
NGC 6366 & \citet{sarajedinietal2007} & HST/ACS F606W/F814W \\
         & \citet{stetsonetal2019} & Ground-based UBVRI \\[+0.05cm]
NGC 6380 & \citet{baumgardtetal2019} & HST/WFC3 F555W/F814W \\
         & \citet{ortolanietal1998} & ESO NTT VI \\[+0.05cm]
NGC 6388 & \citet{sarajedinietal2007} & HST/ACS F606W/F814W \\
         & \citet{stetson2020} & Ground-based UBVRI \\[+0.05cm]
NGC 6397 & \citet{sarajedinietal2007} & HST/ACS F606W/F814W \\
         & \citet{stetson2020} & Ground-based UBVRI \\[+0.05cm]
NGC 6401 & This work  & HST/ACS F606W/F814W \\ 
         & \citet{stetson2020} & Ground-based UBVRI \\[+0.05cm]
NGC 6402 & \citet{piottoetal2002} & HST/WFPC2 F439W/F555W \\
         & \citet{stetson2020} & Ground-based UBVRI \\[+0.05cm]
NGC 6426 & \citet{dotteretal2011} & HST/ACS F606W/F814W \\
         & \citet{hatzidimitriouetal1999} & Ground-based BVRI \\[+0.05cm]
NGC 6440 & \citet{stetson2020} & Ground-based UBVRI \\[+0.05cm]
NGC 6441 & \citet{sarajedinietal2007} & HST/ACS F606W/F814W \\
         & \citet{stetson2020} & Ground-based UBVRI \\[+0.05cm]
NGC 6453 & \citet{baumgardtetal2019} & HST/WFC3 F438W/F555W \\[+0.05cm]
NGC 6496 & \citet{sarajedinietal2007} & HST/ACS F606W/F814W \\
         & \citet{fragaetal2013} & Ground-based BVRI \\[+0.05cm]
NGC 6517 & \citet{baumgardtetal2019} & HST/WFC3 F555W/F814W \\
         & \citet{stetson2020} & Ground-based UBVRI \\[+0.05cm]
NGC 6522 & \citet{kamannetal2018} & HST/WFC3 F555W/F814W \\
         & \citet{stetson2020} & Ground-based UBVRI \\[+0.05cm]
NGC 6528 & \citet{lagioiaetal2014} & HST/ACS F606W/F814W \\[+0.05cm]
         & \citet{stetson2020} & Ground-based UBVRI \\[+0.05cm]
NGC 6535 & \citet{sarajedinietal2007} & HST/ACS F606W/F814W \\
         & \citet{stetson2020} & Ground-based UBVRI \\[+0.05cm]
NGC 6539 & \citet{piottoetal2002} & HST/WFPC2 F439W/F555W \\[+0.05cm]
         & \citet{stetson2020} & Ground-based UBVRI \\[+0.05cm]
NGC 6540 & \citet{piottoetal2002} & HST/WFPC2 F439W/F555W \\
         & \citet{schlaflyetal2018} & Ground-based ugrizY \\[+0.05cm]
NGC 6541 & \citet{sarajedinietal2007} & HST/ACS F606W/F814W \\
         & \citet{stetson2020} & Ground-based UBVRI \\[+0.05cm]
NGC 6544 & \citet{piottoetal2002} & HST/WFPC2 F439W/F555W \\
         & \citet{rosenbergetal2000} & ESO Dutch Telescope VI \\[+0.05cm]
NGC 6553 & \citet{beaulieuetal2001} & HST/WFPC2 F555W/F814W \\
         & \citet{sagaretal1999} & Ground-based VI \\[+0.05cm]
NGC 6558 & \citet{stetson2020} & Ground-based UBVRI \\[+0.05cm]
NGC 6569 & \citet{piottoetal2002} & HST/WFPC2 F439W/F555W \\
         & \citet{stetson2020} & Ground-based UBVRI \\[+0.05cm]
NGC 6584 & \citet{sarajedinietal2007} & HST/ACS F606W/F814W \\[+0.05cm]
         & \citet{stetson2020} & Ground-based UBVRI \\[+0.05cm]
NGC 6624 & \citet{sarajedinietal2007} & HST/ACS F606W/F814W \\
         & \citet{rosenbergetal2000} &  ESO Dutch Telescope VI \\[+0.05cm]
NGC 6626 & \citet{kerberetal2018} & HST/ACS F435W/F625W \\
         & \citet{stetson2020} & Ground-based UBVRI \\[+0.05cm]
NGC 6637 & \citet{sarajedinietal2007} & HST/ACS F606W/F814W \\
         & \citet{stetson2020} & Ground-based UBVRI \\[+0.05cm]
NGC 6638 & \citet{piottoetal2002} & HST/WFPC2 F439W/F555W \\
         & \citet{rosenbergetal2000} &  ESO Dutch Telescope VI \\[+0.05cm]
NGC 6642 & \citet{baumgardtetal2019} & HST/ACS F606W/F814W \\
         & \citet{stetson2020} & Ground-based UBVRI \\[+0.05cm]
NGC 6652 & \citet{sarajedinietal2007} & HST/ACS F606W/F814W \\[+0.05cm]
NGC 6656 & \citet{sarajedinietal2007} & HST/ACS F606W/F814W \\
         & \citet{stetsonetal2019} & Ground-based UBVRI \\[+0.05cm]
NGC 6681 & \citet{sarajedinietal2007} & HST/ACS F606W/F814W \\[+0.05cm]
         & \citet{stetson2020} & Ground-based UBVRI \\[+0.05cm]
NGC 6712 & \citet{piottoetal2002} & HST/WFPC2 F439W/F555W \\
         & \citet{stetsonetal2019} & Ground-based UBVRI \\[+0.05cm]
NGC 6715 & \citet{sarajedinietal2007} & HST/ACS F606W/F814W \\
         & \citet{stetson2020} & Ground-based UBVRI \\[+0.05cm]
\hline\hline
\end{tabular}}
\end{table}

\setcounter{table}{1}

\begin{table}
\caption{continued}
\centering
\resizebox{\columnwidth}{!}{%
\begin{tabular}{@{}lll@{}}
\hline\hline
Cluster & Source of Photometry & Telescope/Instrument/Band \\
\hline\hline
NGC 6717 & \citet{sarajedinietal2007} & HST/ACS F606W/F814W \\
         & \citet{stetson2020} & Ground-based UBVRI \\[+0.05cm]
NGC 6723 & \citet{sarajedinietal2007} & HST/ACS F606W/F814W \\
         & \citet{stetson2020} & Ground-based UBVRI \\[+0.05cm]
NGC 6749 & \citet{stetson2020} & Ground-based UBVRI \\[+0.05cm]
NGC 6752 & \citet{sarajedinietal2007} & HST/ACS F606W/F814W \\
         & \citet{stetsonetal2019} & Ground-based UBVRI \\[+0.05cm]
NGC 6760 & \citet{baumgardtetal2019} & HST/ACS F336W/F625W \\
         & \citet{stetsonetal2019} & Ground-based UBVRI \\[+0.05cm]
NGC 6779 & \citet{sarajedinietal2007} & HST/ACS F606W/F814W \\
         & \citet{stetson2020} & Ground-based UBVRI \\[+0.05cm]
NGC 6809 & \citet{sarajedinietal2007} & HST/ACS F606W/F814W \\
         & \citet{stetsonetal2019} & Ground-based UBVRI \\[+0.05cm]
NGC 6838 & \citet{sarajedinietal2007} & HST/ACS F606W/F814W \\
         & \citet{stetsonetal2019} & Ground-based UBVRI \\[+0.05cm]
NGC 6864 & \citet{baumgardtetal2019} & HST/WFC3 F438W/F555W \\
         & \citet{stetson2020} & Ground-based UBVRI \\[+0.05cm]
NGC 6934 & \citet{sarajedinietal2007} & HST/ACS F606W/F814W \\
         & \citet{stetsonetal2019} & Ground-based UBVRI \\[+0.05cm]
NGC 6981 & \citet{sarajedinietal2007} & HST/ACS F606W/F814W \\
         & \citet{stetsonetal2019} & Ground-based UBVRI \\[+0.05cm]
NGC 7006 & \citet{dotteretal2011} & HST/ACS F606W/F814W \\
         & \citet{stetsonetal2019} & Ground-based UBVRI \\[+0.05cm]
NGC 7078 & \citet{sarajedinietal2007} & HST/ACS F606W/F814W \\
         & \citet{stetsonetal2019} & Ground-based UBVRI \\[+0.05cm]
NGC 7089 & \citet{sarajedinietal2007} & HST/ACS F606W/F814W \\
         & \citet{stetsonetal2019} & Ground-based UBVRI \\[+0.05cm]
NGC 7099 & \citet{sarajedinietal2007} & HST/ACS F606W/F814W \\
         & \citet{stetsonetal2019} & Ground-based UBVRI \\[+0.05cm]
NGC 7492 & \citet{stetsonetal2019} & Ground-based UBVRI \\[+0.05cm]
Pal 1    & \citet{sarajedinietal2007} & HST/ACS F606W/F814W \\
         & \citet{stetson2020} & Ground-based UBVRI \\[+0.05cm]
Pal 2    & \citet{sarajedinietal2007} & HST/ACS F606W/F814W \\
         & \citet{stetson2020} & Ground-based UBVRI \\[+0.05cm]
Pal 3    & \citet{stetsonetal1999} & HST/WFPC2 F555W/F814W \\
         & \citet{stetson2020} & Ground-based UBVRI \\[+0.05cm]
Pal 4    & \citet{franketal2012} & HST/WFPC2 F555W/F814W \\
         & \citet{stetson2020} & Ground-based UBVRI \\[+0.05cm]
Pal 5    & \citet{stetson2020} & Ground-based UBVRI \\[+0.05cm]
Pal 6    & \citet{ortolanietal1995} & Ground-based VI \\
         & \citet{schlaflyetal2018} & Ground-based ugrizY \\[+0.05cm]
Pal 10   & \citet{stetson2020} & Ground-based UBVRI \\[+0.05cm]
Pal 11   & \citet{lewisetal2006} & KPNO/Hiltner V/I  \\[+0.05cm]
         & \citet{stetson2020} & Ground-based UBVRI \\[+0.05cm]
Pal 12   & \citet{sarajedinietal2007} & HST/ACS F606W/F814W \\[+0.05cm]
         & \citet{stetson2020} & Ground-based UBVRI \\[+0.05cm]
Pal 13   & This work & HST/WFC3 F606W/F814W \\[+0.05cm]
         & \citet{stetson2020} & Ground-based UBVRI \\[+0.05cm]
Pal 14   & \citet{franketal2014} & HST/WFPC2 F555W/F814W \\  
         & \citet{stetson2020} & Ground-based UBVRI \\[+0.05cm]
Pal 15   & \citet{dotteretal2011} & HST/ACS F606W/F814W \\[+0.05cm]
         & \citet{stetson2020} & Ground-based UBVRI \\[+0.05cm]
Pyxis    & \citet{dotteretal2011} & HST/ACS F606W/F814W \\[+0.05cm]
         & \citet{stetson2020} & Ground-based UBVRI \\[+0.05cm]
Rup 106  & \citet{dotteretal2011} & HST/ACS F606W/F814W \\
         & \citet{kaluznyetal1995} & Ground-based BV \\[+0.05cm]
Sgr 2    & This work & HST/ACS F606W/F814W  \\
         & \citet{munozetal2018c} & CFHT/MegaCam g/r \\[+0.05cm]
Ter 1    & \citet{schlaflyetal2018} & Ground-based ugrizY \\[+0.05cm]
Ter 2    & \citet{ortolanietal1997}  & Ground-based VI \\[+0.05cm]
Ter 3    & This work & HST/ACS F625W/F658N  \\[+0.05cm]
Ter 4    & \citet{ortolanietal1997b} & Ground-based VI \\[+0.05cm]
Ter 5    & \citet{ortolanietal1996} & Ground-based VI \\[+0.05cm]
\hline\hline
\end{tabular}}
\end{table}

\setcounter{table}{1}

\begin{table}
\caption{continued}
\centering
\resizebox{\columnwidth}{!}{%
\begin{tabular}{@{}lll@{}}
\hline\hline
Cluster & Source of Photometry & Telescope/Instrument/Band \\
\hline\hline
Ter 6    & \citet{barbuyetal1997}  & Ground-based VI \\[+0.05cm]
Ter 7    & \citet{sarajedinietal2007} & HST/ACS F606W/F814W \\
         & \citet{stetson2020} & Ground-based UBVRI \\[+0.05cm]
Ter 8    & \citet{sarajedinietal2007} & HST/ACS F606W/F814W \\
         & \citet{stetson2020} & Ground-based UBVRI \\[+0.05cm]
Ter 9    & \citet{schlaflyetal2018} & Ground-based ugrizY  \\[+0.05cm]
Ter 10   & \citet{ortolanietal2019} & HST/ACS/WFC3 F606W/F160W \\
         & \citet{ortolanietal1997c}  & Ground-based VI \\[+0.05cm]
Ter 12   & \citet{ortolanietal1998}  & Ground-based VI \\[+0.05cm]
Ton 2    & \citet{bicaetal1996} & Ground-based VI \\
         & \citet{schlaflyetal2018} & Ground-based ugrizY \\[+0.05cm]
Whiting 1& \citet{valchevaetal2015} & ESO NTT V/I \\[+0.05cm]
\hline\hline
\end{tabular}}
\end{table}

\section{Derived magnitudes and mass-to-light ratios}

\begin{table*}
\caption{Apparent and absolute $V$-band magnitudes, mass-to-light ratios, and the radii containing 10\% and half the total cluster light in projection together with the surface density brightnesses at these radii for all clusters in this study}
\begin{center}
\begin{tabular}{lcrcrrcc}
\hline
 & \\[-0.3cm]
\multirow{2}{*}{Name} &  $V_{Tot}$ &  $M_V$ &  $M/L_V$ & $r_{10}$ & $r_H$ &  $\mu_{V,10}$ $\;$ & $\mu_{V,H}$ \\ 
 & [mag] & [mag] & [$M_\odot/L_\odot$] & [''] & [''] & [mag/arcsec$^2$] & [mag/arcsec$^2$]\\  
\hline
 & \\[-0.3cm]
AM 1 & 15.07 $\pm$ 0.07 & -6.19 &  0.85 $\pm$ 0.18 &   7.2 &  24.2 & 23.39 & 25.12  \\ 
AM 4 & 16.49 $\pm$ 0.08 & -1.20 &  3.38 $\pm$ 0.73 &  10.5 &  46.1 & 27.27 & 28.60  \\ 
Arp 2 & 11.65 $\pm$ 0.07 & -5.94 &  1.91 $\pm$ 0.35 &  33.5 & 111.7 & 22.89 & 24.81  \\ 
BH 140 & $\;$ 9.14 $\pm$ 0.11 & -6.21 &  1.88 $\pm$ 0.31 &  87.1 & 263.7 & 22.73 & 24.16 \\ 
BH 261 & 10.75 $\pm$ 0.15 & -4.43 &  1.89 $\pm$ 0.32 &  19.2 &  92.9 & 21.76 & 25.04  \\ 
Crater & 15.74 $\pm$ 0.06 & -5.07 &  1.18 $\pm$ 0.24 &   9.1 &  27.2 & 24.12 & 25.98  \\ 
Djor 1 & 13.08 $\pm$ 0.22 & -6.66 &  1.88 $\pm$ 0.56 &  19.7 &  86.8 & 23.41 & 25.02  \\ 
Djor 2 & 10.70 $\pm$ 0.19 & -6.52 &  2.29 $\pm$ 0.45 &  18.0 &  71.3 & 21.36 & 22.60  \\ 
E 3 & 11.84 $\pm$ 0.05 & -3.62 &  1.20 $\pm$ 0.24 &  31.8 & 111.2 & 23.72 & 24.68  \\ 
Eridanus & 15.02 $\pm$ 0.06 & -5.56 &  0.67 $\pm$ 0.14 &  10.1 &  33.6 & 24.02 & 25.53  \\ 
ESO 280 & 12.51 $\pm$ 0.09 & -4.28 &  1.69 $\pm$ 0.44 &  17.6 &  69.6 & 22.45 & 24.70  \\ 
ESO 452 & 11.77 $\pm$ 0.09 & -3.82 &  2.70 $\pm$ 1.43 &  18.2 &  67.5 & 21.05 & 24.83  \\ 
FSR 1716 & 13.05 $\pm$ 0.16 & -4.82 &  2.18 $\pm$ 2.53 &  37.2 & 120.0 & 24.76 & 26.07  \\ 
FSR 1735 & 14.38 $\pm$ 0.09 & -7.09 &  1.33 $\pm$ 0.29 &  11.8 &  44.4 & 24.08 & 25.50  \\ 
FSR 1758 & $\;$ 9.14 $\pm$ 0.11 & -8.96 &  1.79 $\pm$ 0.31 &  88.3 & 235.0 & 22.37 & 23.63 \\ 
HP 1 & 11.07 $\pm$ 0.19 & -6.56 &  2.91 $\pm$ 0.90 &  21.9 &  75.9 & 21.74 & 23.44  \\ 
IC 1257 & 13.81 $\pm$ 0.18 & -5.44 &  1.55 $\pm$ 0.42 &   5.6 &  30.2 & 20.91 & 24.35  \\ 
IC 1276 & $\;$ 9.93 $\pm$ 0.09 & -7.08 &  1.43 $\pm$ 0.41 &  43.3 & 160.3 & 22.15 & 24.08 \\ 
IC 4499 & $\;$ 9.84 $\pm$ 0.08 & -7.18 &  2.03 $\pm$ 0.42 &  31.8 & 110.5 & 21.22 & 22.85 \\ 
Lil 1 & 15.73 $\pm$ 0.09 & -9.04 &  2.46 $\pm$ 0.45 &   5.1 &  26.2 & 23.26 & 26.16  \\ 
Lynga 7 & $\;$ 9.87 $\pm$ 0.16 & -6.91 &  2.06 $\pm$ 0.55 &  34.5 & 107.6 & 21.49 & 22.40 \\ 
NGC 104 & $\;$ 4.08 $\pm$ 0.08 & -9.28 &  1.89 $\pm$ 0.14 &  26.8 & 168.5 & 15.40 & 18.79 \\ 
NGC 288 & $\;$ 8.09 $\pm$ 0.07 & -6.77 &  2.14 $\pm$ 0.15 &  46.9 & 143.6 & 20.38 & 21.69 \\ 
NGC 362 & $\;$ 6.48 $\pm$ 0.06 & -8.49 &  1.46 $\pm$ 0.08 &  10.9 &  51.1 & 15.75 & 18.40 \\ 
NGC 1261 & $\;$ 8.31 $\pm$ 0.06 & -7.77 &  1.62 $\pm$ 0.10 &  10.7 &  41.1 & 17.61 & 19.25 \\ 
NGC 1851 & $\;$ 7.07 $\pm$ 0.07 & -8.26 &  1.63 $\pm$ 0.11 &   5.1 &  30.6 & 15.06 & 17.74 \\ 
NGC 1904 & $\;$ 7.94 $\pm$ 0.07 & -7.65 &  1.47 $\pm$ 0.15 &   7.8 &  40.1 & 16.69 & 19.11 \\ 
NGC 2298 & $\;$ 9.06 $\pm$ 0.05 & -6.54 &  1.63 $\pm$ 0.34 &  12.6 &  50.3 & 18.69 & 20.99 \\ 
NGC 2419 & 10.56 $\pm$ 0.07 & -9.29 &  2.05 $\pm$ 0.34 &  12.3 &  46.8 & 20.11 & 22.00  \\ 
NGC 2808 & $\;$ 6.14 $\pm$ 0.06 & -9.59 &  1.52 $\pm$ 0.09 &  11.1 &  50.6 & 15.58 & 17.73 \\ 
NGC 3201 & $\;$ 6.77 $\pm$ 0.07 & -7.29 &  2.01 $\pm$ 0.13 &  39.2 & 166.4 & 19.40 & 21.11 \\ 
NGC 4147 & 10.29 $\pm$ 0.05 & -6.07 &  1.66 $\pm$ 0.41 &   5.4 &  28.6 & 18.69 & 21.02  \\ 
NGC 4372 & $\;$ 7.37 $\pm$ 0.07 & -7.64 &  2.10 $\pm$ 0.18 &  68.4 & 188.7 & 20.57 & 21.65 \\ 
NGC 4590 & $\;$ 8.00 $\pm$ 0.07 & -7.19 &  2.01 $\pm$ 0.22 &  24.2 &  89.0 & 19.01 & 20.85 \\ 
NGC 4833 & $\;$ 7.19 $\pm$ 0.05 & -7.89 &  1.36 $\pm$ 0.11 &  30.2 & 104.5 & 18.51 & 20.23 \\ 
NGC 5024 & $\;$ 7.71 $\pm$ 0.06 & -8.62 &  1.74 $\pm$ 0.17 &  17.0 &  72.6 & 17.68 & 20.06 \\ 
NGC 5053 & $\;$ 9.93 $\pm$ 0.06 & -6.28 &  2.60 $\pm$ 0.59 &  53.2 & 146.0 & 22.02 & 23.21 \\ 
NGC 5139 & $\;$ 3.50 $\pm$ 0.06 & -10.47 &  2.46 $\pm$ 0.15 &  90.0 & 284.9 & 17.23 & 18.58 \\ 
NGC 5272 & $\;$ 6.38 $\pm$ 0.06 & -8.56 &  1.64 $\pm$ 0.12 &  16.3 &  69.4 & 16.64 & 18.71 \\ 
NGC 5286 & $\;$ 7.35 $\pm$ 0.06 & -8.69 &  1.42 $\pm$ 0.09 &  10.2 &  44.8 & 16.25 & 18.78 \\ 
NGC 5466 & $\;$ 9.32 $\pm$ 0.06 & -6.70 &  1.44 $\pm$ 0.30 &  41.2 & 123.1 & 21.46 & 22.26 \\ 
NGC 5634 & $\;$ 9.51 $\pm$ 0.11 & -7.82 &  1.91 $\pm$ 0.46 &   6.4 &  36.1 & 18.16 & 20.73 \\ 
NGC 5694 & $\;$ 9.87 $\pm$ 0.18 & -8.27 &  2.08 $\pm$ 0.42 &   2.6 &  18.6 & 16.56 & 19.47 \\ 
NGC 5824 & $\;$ 8.86 $\pm$ 0.10 & -9.05 &  2.17 $\pm$ 0.24 &   5.0 &  29.8 & 16.38 & 19.74 \\ 
NGC 5897 & $\;$ 8.48 $\pm$ 0.06 & -7.30 &  2.19 $\pm$ 0.29 &  43.0 & 126.0 & 20.81 & 21.66 \\ 
NGC 5904 & $\;$ 5.95 $\pm$ 0.05 & -8.54 &  1.62 $\pm$ 0.08 &  23.3 &  97.2 & 16.79 & 19.06 \\ 
NGC 5927 & $\;$ 7.74 $\pm$ 0.07 & -8.45 &  1.58 $\pm$ 0.10 &  23.5 &  86.4 & 18.52 & 20.67 \\ 
NGC 5946 & $\;$ 9.50 $\pm$ 0.13 & -7.31 &  1.89 $\pm$ 0.41 &   9.0 &  41.4 & 18.53 & 21.00 \\ 
NGC 5986 & $\;$ 7.71 $\pm$ 0.06 & -8.28 &  1.89 $\pm$ 0.18 &  15.1 &  54.9 & 17.95 & 19.48 \\ 
\hline
\end{tabular}
\end{center}
\end{table*}
 
\addtocounter{table}{-1}
\begin{table*}
\caption{continued}
\begin{center}
\begin{tabular}{lcrcrrcc}
\hline
 & \\[-0.3cm]
\multirow{2}{*}{Name} &  $V_{Tot}$ &  $M_V$ &  $M/L_V$ & $r_{10}$ & $r_H$ &  $\mu_{V,10}$ $\;$ & $\mu_{V,H}$ \\ 
 & [mag] & [mag] & [$M_\odot/L_\odot$] & [''] & [''] & [mag/arcsec$^2$] & [mag/arcsec$^2$]\\  
\hline
 & \\[-0.3cm]
NGC 6093 & $\;$ 7.42 $\pm$ 0.05 & -8.13 &  1.93 $\pm$ 0.12 &   7.8 &  35.4 & 15.99 & 18.39 \\ 
NGC 6101 & $\;$ 8.68 $\pm$ 0.05 & -7.02 &  2.35 $\pm$ 0.66 &  43.2 & 138.1 & 20.64 & 22.12 \\ 
NGC 6121 & $\;$ 5.66 $\pm$ 0.07 & -6.86 &  1.97 $\pm$ 0.13 &  59.9 & 279.4 & 18.54 & 20.97 \\ 
NGC 6139 & $\;$ 8.97 $\pm$ 0.12 & -8.31 &  1.92 $\pm$ 0.37 &   7.9 &  41.8 & 18.04 & 20.34 \\ 
NGC 6144 & $\;$ 9.25 $\pm$ 0.06 & -6.62 &  1.66 $\pm$ 0.46 &  28.0 &  92.2 & 20.63 & 22.29 \\ 
NGC 6171 & $\;$ 8.28 $\pm$ 0.06 & -6.63 &  2.11 $\pm$ 0.17 &  28.0 & 104.9 & 19.17 & 21.41 \\ 
NGC 6205 & $\;$ 5.81 $\pm$ 0.07 & -8.40 &  2.32 $\pm$ 0.18 &  28.8 &  97.5 & 17.02 & 18.67 \\ 
NGC 6218 & $\;$ 7.08 $\pm$ 0.07 & -6.85 &  1.85 $\pm$ 0.14 &  32.2 & 115.4 & 18.81 & 20.15 \\ 
NGC 6229 & $\;$ 9.33 $\pm$ 0.08 & -8.13 &  1.89 $\pm$ 0.63 &   5.6 &  21.6 & 17.05 & 19.23 \\ 
NGC 6235 & $\;$ 9.62 $\pm$ 0.07 & -6.99 &  2.10 $\pm$ 0.68 &  14.9 &  57.5 & 19.40 & 21.44 \\ 
NGC 6254 & $\;$ 6.62 $\pm$ 0.07 & -7.73 &  1.80 $\pm$ 0.12 &  31.4 & 123.0 & 17.91 & 20.20 \\ 
NGC 6256 & 10.55 $\pm$ 0.16 & -6.85 &  2.43 $\pm$ 0.85 &  15.1 &  79.3 & 20.93 & 23.36  \\ 
NGC 6266 & $\;$ 6.60 $\pm$ 0.09 & -8.89 &  2.00 $\pm$ 0.18 &  12.4 &  58.4 & 16.44 & 18.74 \\ 
NGC 6273 & $\;$ 6.88 $\pm$ 0.07 & -8.89 &  2.15 $\pm$ 0.17 &  19.5 &  78.3 & 17.42 & 19.35 \\ 
NGC 6284 & $\;$ 9.31 $\pm$ 0.14 & -7.46 &  1.53 $\pm$ 0.39 &   8.1 &  43.6 & 18.31 & 20.97 \\ 
NGC 6287 & $\;$ 9.27 $\pm$ 0.10 & -7.46 &  1.88 $\pm$ 0.44 &  11.0 &  50.0 & 19.12 & 20.72 \\ 
NGC 6293 & $\;$ 8.40 $\pm$ 0.16 & -7.41 &  1.75 $\pm$ 0.31 &  19.3 &  53.5 & 17.97 & 20.28 \\ 
NGC 6304 & $\;$ 8.18 $\pm$ 0.10 & -7.30 &  1.94 $\pm$ 0.25 &  13.1 &  66.9 & 17.89 & 20.63 \\ 
NGC 6316 & $\;$ 9.02 $\pm$ 0.26 & -7.98 &  2.17 $\pm$ 0.71 &   9.8 &  55.2 & 18.02 & 21.33 \\ 
NGC 6325 & 10.89 $\pm$ 0.10 & -6.39 &  2.78 $\pm$ 0.53 &   9.7 &  46.9 & 19.66 & 22.31  \\ 
NGC 6333 & $\;$ 7.65 $\pm$ 0.07 & -8.15 &  2.08 $\pm$ 0.21 &  17.6 &  68.4 & 17.90 & 19.93 \\ 
NGC 6341 & $\;$ 6.50 $\pm$ 0.05 & -8.19 &  1.92 $\pm$ 0.10 &  14.0 &  58.2 & 16.35 & 18.30 \\ 
NGC 6342 & $\;$ 9.86 $\pm$ 0.14 & -6.19 &  2.47 $\pm$ 0.55 &   7.9 &  38.7 & 18.73 & 21.35 \\ 
NGC 6352 & $\;$ 8.06 $\pm$ 0.07 & -6.24 &  2.14 $\pm$ 0.17 &  27.0 & 109.6 & 19.37 & 21.35 \\ 
NGC 6355 & 10.04 $\pm$ 0.11 & -7.17 &  1.91 $\pm$ 0.35 &  11.4 &  56.2 & 19.30 & 22.12  \\ 
NGC 6356 & $\;$ 8.32 $\pm$ 0.10 & -8.45 &  2.13 $\pm$ 0.50 &  11.8 &  51.6 & 19.08 & 20.51 \\ 
NGC 6362 & $\;$ 7.45 $\pm$ 0.07 & -7.16 &  1.87 $\pm$ 0.14 &  44.3 & 139.2 & 19.21 & 21.10 \\ 
NGC 6366 & $\;$ 8.85 $\pm$ 0.07 & -6.24 &  1.51 $\pm$ 0.19 &  67.6 & 218.8 & 21.78 & 23.42 \\ 
NGC 6380 & 10.70 $\pm$ 0.18 & -7.88 &  1.97 $\pm$ 0.50 &  16.4 &  68.4 & 20.83 & 23.46  \\ 
NGC 6388 & $\;$ 6.81 $\pm$ 0.06 & -9.49 &  2.15 $\pm$ 0.12 &   7.9 &  49.6 & 15.70 & 18.73 \\ 
NGC 6397 & $\;$ 5.45 $\pm$ 0.06 & -7.05 &  1.58 $\pm$ 0.10 &  30.2 & 181.7 & 17.79 & 20.37 \\ 
NGC 6401 & $\;$ 9.91 $\pm$ 0.13 & -6.76 &  3.19 $\pm$ 1.69 &  13.4 &  57.9 & 19.46 & 21.74 \\ 
NGC 6402 & $\;$ 7.87 $\pm$ 0.09 & -8.84 &  1.91 $\pm$ 0.24 &  25.7 &  72.1 & 18.82 & 20.11 \\ 
NGC 6426 & 11.13 $\pm$ 0.07 & -6.47 &  1.95 $\pm$ 0.81 &  15.6 &  52.4 & 21.01 & 22.79  \\ 
NGC 6440 & $\;$ 8.98 $\pm$ 0.19 & -8.92 &  1.58 $\pm$ 0.36 &   6.9 &  32.2 & 17.60 & 19.81 \\ 
NGC 6441 & $\;$ 7.12 $\pm$ 0.12 & -9.70 &  1.94 $\pm$ 0.24 &   6.9 &  35.0 & 15.87 & 18.29 \\ 
NGC 6453 & $\;$ 9.19 $\pm$ 0.19 & -8.12 &  1.39 $\pm$ 0.34 &  10.7 &  71.1 & 18.56 & 21.75 \\ 
NGC 6496 & $\;$ 8.64 $\pm$ 0.14 & -6.62 &  1.70 $\pm$ 0.34 &  30.1 &  93.3 & 20.29 & 21.59 \\ 
NGC 6517 & 10.70 $\pm$ 0.09 & -7.78 &  2.27 $\pm$ 0.67 &   6.8 &  37.9 & 19.43 & 21.98  \\ 
NGC 6522 & $\;$ 8.14 $\pm$ 0.07 & -7.86 &  1.97 $\pm$ 0.18 &  14.0 &  71.0 & 17.79 & 20.99 \\ 
NGC 6528 & $\;$ 9.71 $\pm$ 0.15 & -6.32 &  1.69 $\pm$ 0.31 &   7.8 &  50.0 & 18.51 & 21.49 \\ 
NGC 6535 & 10.10 $\pm$ 0.06 & -5.02 &  1.52 $\pm$ 0.25 &  16.4 &  86.9 & 20.79 & 23.46  \\ 
NGC 6539 & $\;$ 9.95 $\pm$ 0.07 & -7.69 &  2.15 $\pm$ 0.29 &  25.4 &  93.9 & 21.09 & 22.69 \\ 
NGC 6540 & $\;$ 9.74 $\pm$ 0.17 & -5.89 &  1.97 $\pm$ 0.70 &  10.6 &  70.9 & 20.09 & 22.57 \\ 
NGC 6541 & $\;$ 6.62 $\pm$ 0.06 & -8.32 &  1.38 $\pm$ 0.09 &  10.1 &  63.1 & 16.16 & 19.12 \\ 
NGC 6544 & $\;$ 7.86 $\pm$ 0.32 & -6.58 &  2.33 $\pm$ 0.84 &  21.1 & 124.7 & 19.24 & 21.36 \\ 
NGC 6553 & $\;$ 8.04 $\pm$ 0.09 & -7.80 &  2.50 $\pm$ 0.26 &  21.8 &  88.2 & 18.78 & 20.91 \\ 
NGC 6558 & $\;$ 9.66 $\pm$ 0.17 & -6.00 &  1.82 $\pm$ 0.50 &  11.7 &  45.1 & 18.94 & 21.22 \\ 
NGC 6569 & $\;$ 8.89 $\pm$ 0.12 & -7.75 &  1.99 $\pm$ 0.30 &  14.0 &  50.8 & 18.69 & 20.58 \\ 
NGC 6584 & $\;$ 8.76 $\pm$ 0.05 & -7.30 &  1.14 $\pm$ 0.38 &  14.8 &  52.5 & 18.60 & 20.23 \\ 
NGC 6624 & $\;$ 8.04 $\pm$ 0.11 & -7.18 &  1.50 $\pm$ 0.16 &   9.3 &  58.7 & 17.18 & 20.23 \\ 
\hline
\end{tabular}
\end{center}
\end{table*}
 
\addtocounter{table}{-1}
\begin{table*}
\caption{continued}
\begin{center}
\begin{tabular}{lcrcrrcc}
\hline
 & \\[-0.3cm]
\multirow{2}{*}{Name} &  $V_{Tot}$ &  $M_V$ &  $M/L_V$ & $r_{10}$ & $r_H$ &  $\mu_{V,10}$ $\;$ & $\mu_{V,H}$ \\ 
 & [mag] & [mag] & [$M_\odot/L_\odot$] & [''] & [''] & [mag/arcsec$^2$] & [mag/arcsec$^2$]\\  
\hline
 & \\[-0.3cm]
NGC 6626 & $\;$ 6.85 $\pm$ 0.10 & -8.06 &  2.01 $\pm$ 0.21 &  13.1 &  61.9 & 16.68 & 18.99 \\ 
NGC 6637 & $\;$ 7.62 $\pm$ 0.06 & -7.54 &  1.83 $\pm$ 0.22 &  14.0 &  55.3 & 17.22 & 19.56 \\ 
NGC 6638 & $\;$ 8.79 $\pm$ 0.15 & -7.55 &  1.34 $\pm$ 0.35 &   9.4 &  39.5 & 17.45 & 20.02 \\ 
NGC 6642 & $\;$ 9.65 $\pm$ 0.13 & -6.30 &  2.27 $\pm$ 0.53 &   6.9 &  34.8 & 18.04 & 20.79 \\ 
NGC 6652 & $\;$ 8.92 $\pm$ 0.06 & -6.36 &  1.72 $\pm$ 0.31 &   6.2 &  32.0 & 17.13 & 19.48 \\ 
NGC 6656 & $\;$ 5.06 $\pm$ 0.07 & -8.54 &  1.83 $\pm$ 0.13 &  47.0 & 198.9 & 17.63 & 19.65 \\ 
NGC 6681 & $\;$ 7.91 $\pm$ 0.06 & -7.16 &  1.84 $\pm$ 0.11 &   8.0 &  47.5 & 17.13 & 19.56 \\ 
NGC 6712 & $\;$ 8.59 $\pm$ 0.07 & -7.01 &  1.71 $\pm$ 0.16 &  20.6 &  71.4 & 19.51 & 20.56 \\ 
NGC 6715 & $\;$ 7.57 $\pm$ 0.10 & -9.81 &  2.06 $\pm$ 0.20 &   4.7 &  28.3 & 15.38 & 18.26 \\ 
NGC 6717 & $\;$ 8.98 $\pm$ 0.06 & -5.59 &  1.70 $\pm$ 0.40 &  17.1 & 109.6 & 18.90 & 22.63 \\ 
NGC 6723 & $\;$ 7.21 $\pm$ 0.06 & -7.54 &  1.96 $\pm$ 0.17 &  26.5 &  88.2 & 18.07 & 19.84 \\ 
NGC 6749 & 10.90 $\pm$ 0.10 & -8.21 &  1.34 $\pm$ 0.60 &  30.9 & 131.4 & 22.94 & 24.74  \\ 
NGC 6752 & $\;$ 5.34 $\pm$ 0.08 & -7.92 &  2.09 $\pm$ 0.17 &  24.2 & 146.1 & 16.56 & 19.41 \\ 
NGC 6760 & $\;$ 8.89 $\pm$ 0.08 & -8.00 &  1.89 $\pm$ 0.27 &  17.7 &  79.4 & 19.07 & 21.72 \\ 
NGC 6779 & $\;$ 8.15 $\pm$ 0.08 & -7.59 &  1.67 $\pm$ 0.22 &  16.1 &  59.1 & 18.28 & 20.09 \\ 
NGC 6809 & $\;$ 6.29 $\pm$ 0.06 & -7.58 &  2.07 $\pm$ 0.13 &  56.0 & 177.7 & 18.79 & 20.12 \\ 
NGC 6838 & $\;$ 7.16 $\pm$ 0.05 & -6.62 &  1.35 $\pm$ 0.09 &  44.0 & 173.1 & 19.52 & 21.74 \\ 
NGC 6864 & $\;$ 8.50 $\pm$ 0.06 & -8.66 &  1.63 $\pm$ 0.34 &   4.2 &  20.9 & 15.89 & 18.39 \\ 
NGC 6934 & $\;$ 8.75 $\pm$ 0.06 & -7.50 &  1.77 $\pm$ 0.30 &  10.4 &  39.0 & 17.58 & 19.90 \\ 
NGC 6981 & $\;$ 9.32 $\pm$ 0.06 & -6.99 &  1.60 $\pm$ 0.31 &  16.0 &  51.7 & 18.88 & 20.66 \\ 
NGC 7006 & 10.69 $\pm$ 0.06 & -7.54 &  1.52 $\pm$ 0.40 &   6.0 &  23.0 & 18.48 & 20.48  \\ 
NGC 7078 & $\;$ 6.29 $\pm$ 0.10 & -9.07 &  1.55 $\pm$ 0.15 &   5.3 &  39.6 & 14.75 & 17.74 \\ 
NGC 7089 & $\;$ 6.47 $\pm$ 0.06 & -8.82 &  1.78 $\pm$ 0.11 &  11.8 &  48.9 & 16.15 & 18.26 \\ 
NGC 7099 & $\;$ 7.37 $\pm$ 0.07 & -7.24 &  2.04 $\pm$ 0.17 &   9.4 &  62.7 & 17.00 & 19.65 \\ 
NGC 7492 & 11.14 $\pm$ 0.05 & -5.98 &  1.40 $\pm$ 0.41 &  21.3 &  64.4 & 22.09 & 22.89  \\ 
Pal 1 & 13.95 $\pm$ 0.05 & -1.72 &  2.45 $\pm$ 0.49 &   8.1 &  33.6 & 22.92 & 25.46  \\ 
Pal 2 & 12.64 $\pm$ 0.09 & -8.38 &  1.22 $\pm$ 0.53 &   9.2 &  38.5 & 21.46 & 23.62  \\ 
Pal 3 & 14.52 $\pm$ 0.06 & -5.44 &  1.08 $\pm$ 0.20 &  15.1 &  43.9 & 24.64 & 25.49  \\ 
Pal 4 & 14.23 $\pm$ 0.07 & -5.90 &  1.41 $\pm$ 0.61 &  12.4 &  35.0 & 23.49 & 24.65  \\ 
Pal 5 & 11.86 $\pm$ 0.07 & -4.90 &  1.90 $\pm$ 0.34 &  68.4 & 192.8 & 25.29 & 26.16  \\ 
Pal 6 & 11.60 $\pm$ 0.13 & -6.74 &  2.29 $\pm$ 0.77 &  20.1 &  78.9 & 21.76 & 24.03  \\ 
Pal 10 & 12.37 $\pm$ 0.05 & -6.63 &  1.72 $\pm$ 0.95 &  26.9 &  98.4 & 23.29 & 25.19  \\ 
Pal 11 & 11.86 $\pm$ 0.05 & -5.16 &  1.13 $\pm$ 0.49 &  28.8 &  85.4 & 22.86 & 24.05  \\ 
Pal 12 & 11.99 $\pm$ 0.07 & -4.47 &  1.21 $\pm$ 0.25 &  23.1 &  76.5 & 23.10 & 24.79  \\ 
Pal 13 & 13.89 $\pm$ 0.08 & -3.27 &  1.74 $\pm$ 0.35 &  20.5 & 113.4 & 24.52 & 27.90  \\ 
Pal 14 & 14.13 $\pm$ 0.06 & -5.22 &  1.74 $\pm$ 0.34 &  23.5 &  68.6 & 24.62 & 26.50  \\ 
Pal 15 & 13.55 $\pm$ 0.08 & -5.77 &  2.94 $\pm$ 0.62 &  32.9 &  93.2 & 24.67 & 25.97  \\ 
Pyxis & 13.21 $\pm$ 0.07 & -5.64 &  1.71 $\pm$ 0.31 &  34.8 &  96.4 & 24.31 & 26.43  \\ 
Rup 106 & 11.05 $\pm$ 0.05 & -6.20 &  1.34 $\pm$ 0.24 &  25.8 &  76.2 & 22.44 & 22.90  \\ 
Sgr 2 & 14.04 $\pm$ 0.12 & -5.62 &  1.35 $\pm$ 0.29 &  30.5 &  90.5 & 24.23 & 26.30  \\ 
Ter 1 & 12.41 $\pm$ 0.11 & -7.89 &  1.70 $\pm$ 0.34 &  13.0 &  52.6 & 22.20 & 24.16  \\ 
Ter 2 & 13.11 $\pm$ 0.15 & -7.06 &  1.92 $\pm$ 1.07 &  12.6 &  67.6 & 22.89 & 25.27  \\ 
Ter 3 & 10.59 $\pm$ 0.21 & -6.22 &  2.15 $\pm$ 0.89 &  40.6 & 124.9 & 23.36 & 99.99  \\ 
Ter 4 & 13.43 $\pm$ 0.39 & -7.06 &  2.32 $\pm$ 1.09 &  34.4 & 156.2 & 25.31 & 99.99  \\ 
Ter 5 & 12.36 $\pm$ 0.12 & -8.72 &  3.15 $\pm$ 0.41 &  10.8 &  52.7 & 21.65 & 24.17  \\ 
Ter 6 & 14.47 $\pm$ 0.10 & -6.95 &  2.29 $\pm$ 0.98 &   7.2 &  50.6 & 22.94 & 26.40  \\ 
Ter 7 & 11.86 $\pm$ 0.07 & -5.15 &  1.98 $\pm$ 0.37 &  15.8 &  54.1 & 20.71 & 23.63  \\ 
Ter 8 & 11.04 $\pm$ 0.05 & -6.46 &  1.79 $\pm$ 0.41 &  38.1 & 114.4 & 22.90 & 23.78  \\ 
Ter 9 & 12.73 $\pm$ 0.12 & -6.56 &  2.30 $\pm$ 0.42 &  11.8 &  57.8 & 21.78 & 25.11  \\ 
Ter 10 & 14.73 $\pm$ 0.16 & -7.07 &  5.27 $\pm$ 1.34 &  10.7 &  69.5 & 24.09 & 27.10  \\ 
Ter 12 & 13.82 $\pm$ 0.17 & -5.88 &  3.13 $\pm$ 0.79 &  18.0 &  74.5 & 24.22 & 26.18  \\ 
\hline
\end{tabular}
\end{center}
\end{table*}
 
\addtocounter{table}{-1}
\begin{table*}
\caption{continued}
\begin{center}
\begin{tabular}{lcrcrrcc}
\hline
 & \\[-0.3cm]
\multirow{2}{*}{Name} &  $V_{Tot}$ &  $M_V$ &  $M/L_V$ & $r_{10}$ & $r_H$ &  $\mu_{V,10}$ $\;$ & $\mu_{V,H}$ \\ 
 & [mag] & [mag] & [$M_\odot/L_\odot$] & [''] & [''] & [mag/arcsec$^2$] & [mag/arcsec$^2$]\\  
\hline
 & \\[-0.3cm]
Ton 2 & 11.66 $\pm$ 0.11 & -6.21 &  2.83 $\pm$ 1.17 &  24.2 &  85.6 & 22.44 & 24.20  \\ 
Whiting 1 & 14.61 $\pm$ 0.12 & -4.23 &  0.43 $\pm$ 0.10 &   9.8 &  55.2 & 23.01 & 26.80  \\ 
\hline
\end{tabular}
\end{center}
\label{magtab}
\end{table*}

\end{appendix}

\bibliographystyle{pasa-mnras}
\bibliography{mybib}

\end{document}